\documentclass[english]{article}
\usepackage{graphicx}
\usepackage{epsfig}
\usepackage{authblk}
\usepackage{subfigure}
\usepackage{wrapfig}
\usepackage[T1]{fontenc}
\usepackage[latin9]{inputenc}
\usepackage{units}
\usepackage{amsmath}
\usepackage{amssymb}
\usepackage{esint}
\usepackage{babel}
\usepackage{color}
\usepackage{float}
\usepackage{lipsum}  
\title{ {\small \it under consideration for publication in Physics of Fluids} \\ \vspace{8mm}
Investigation of wakes generated by fractal plates in the compressible flow regime using large-eddy simulations}

\author{Omar Es-Sahli}%
\author{Adrian Sescu}%
\affil[1]{Aerospace Engineering, Mississippi State University, Starkville, MS 39762, USA}%
\author{Mohammed Z. Afsar}%
\affil[2]{Mechanical \& Aerospace Engineering, University of Strathclyde,  Glasgow, G1 1XJ, UK}%
\author{Oliver R.H. Buxton\thanks}%
\affil[3]{Department of Aeronautics, Imperial College London, S Kensington Campus, SW7 2AZ, UK}%

\begin{document}

\date{}

\maketitle

\begin{abstract}

{\small
We investigate flows interacting with a square and a fractal shape multi-scale structures in the compressible regime for Mach numbers at subsonic and supersonic upstream conditions using large-eddy-simulations (LES). We also aim at identifying similarities and differences that these interactions have with corresponding interactions in the canonical incompressible flow problem. To account for the geometrical complexity associated with the fractal structures, we apply an immersed boundary method to model the no-slip boundary condition at the solid surfaces, with adequate mesh resolution in the vicinity of the small fractal features. We validate the numerical results through extensive comparisons with experimental wind tunnel measurements at a low Mach number. Similar to the incompressible flow case results, we find a break-up of the flow structures by the fractal plate and an increase in turbulent mixing in the downstream direction. As the Mach number increases, we observe noticeable wake meandering and higher spread rate of the wake in the lateral direction perpendicular to the streamwise-spanwise plane. Although not significant, we quantify the difference between the square and the fractal plates using two-point velocity correlations across the Mach number range. The wakes generated by the fractal plate in the compressible regime showed lower turbulent kinetic energy (TKE) and energy spectra levels compared to those of the square case. Moreover, results in terms of the near-field pressure spectra seem to indicate that the fractal plate has the potential to reduce the aerodynamic noise.}

\end{abstract}

\maketitle

\section{Introduction}\label{}

A fractal -- a concept invented by the mathematician Benoit Mandelbrot in 1975 -- represents a detailed, recursive, and infinitely self-similar mathematical set that exhibits similar patterns at increasingly smaller length scales. In other words, fractals are objects that display self-similarity over a wide range of scales. Soon after its introduction, the generic mathematical concept was applied to fluid mechanics research, first, as a means to describe the phenomenological features of flows, e.g the turbulent/non-turbulent interface, and later on, to manipulate turbulent fluid flows. Studies in the past focused mainly on the low Mach number (incompressible) flows and intensively sought the benefits that these multiscale structures can have on engineering applications. 

The application of fractals to turbulent flows traces back to the 1980s, when Sreenivasan and Meneveau \cite{Sreenivasan}  experimentally investigated the applicability of fractals to turbulence as previously speculated by Mandelbrot \cite{Mandelbrot1,Mandelbrot2} (see also Meneveau and Sreenivasan \cite{Meneveau_and_Sreenivasan}). Their work shed light on the subject through carefully designed experiments, subsequently leading to several other studies conducted to assess the numerous applications of fractals (see, for example, Vassilicos and Hunt \cite{Vassilicos}, Procaccia et al. \cite{Procaccia}, Scotti et al. \cite{Scotti}).  

While such early studies focused mainly on a possible fractal-like representation of turbulence, research concerning an upstream uniform flow interacting with a fractal geometry has received new consideration more recently. This direction of research has opened the door to various novel applications, such as fractal-grid-generated turbulence which has proved popular in the research community; see for example, Mazzi and Vassilicos \cite{Mazzi}, Hurst and Vassilicos \cite{Hurst}, Seoud and Vassilicos \cite{Seoud}, Mazellier and Vassilicos \cite{Mazellier}, Stressing et al. \cite{Stresing}, and Suzuki at al \cite{Suzuki_1,Suzuki_2,Suzuki_3}. Their results in terms of fractal-grid-generated turbulence compared to turbulence generated by regular grids with the same blockage ratio, uncovered distinctive characteristics such as a significant increase in turbulence intensities, high local turbulent Reynolds numbers $R_L$, increased decay rates of TKE, or a constant ratio between the integral turbulent scale and the Taylor micro-scale. Other studies found that, for the same mesh Raynolds number, fractal grids remarkably enhanced the level of turbulent mixing compared to that observed in regular grid turbulence (Suzuki at al. \cite{Suzuki_1}). These studies considered different types of fractal-generating patterns to create fractal grids. In the same context of fractal-generated turbulence, several researchers adopted a mix between the classical and fractal grids, and employed active grids that allow dynamical generation of flow fields. Experiments of this kind showed strong differences in the turbulent decay \cite{Weitemeyer} and allowed elevated Reynolds numbers further downstream compared to passive grids \cite{Thormann}. See also Geipel et al. \cite{Geipel}, Laizet et al. \cite{Laizet2011,Laizet2012}, and Nagata et al. \cite{Nagata2013} for more work on fractal-generated turbulence.

Another research direction sought to understand the effect that fractal geometries have on flows such as wakes generated by fractal plates. In the study of Nedi$\acute{c}$ et al. \cite{Nedic2013}, it was observed that the statistics of a fractal plate generated wake can be considered axisymmetric to a good approximation for $\it{x/l \geq 10}$, where $\it{l}$ is the characteristic length of the plates (see also Nedi$\acute{c}$ \cite{Nedic1} or Nedi$\acute{c}$ et al. \cite{Nedic2}). Forest canopies characterized by fractal/tree-like objects were experimentally studied by Bai et al. \cite{Bai2012,Bai2013,Bai2015}, whose results suggest that models of the mixing length for flows through single or sparse canopies of multi-scale trees must incorporate information about multi-scale clustering of branches. Schr\"{o}ttle and D\"{o}rnbrack \cite{Schrottle} focused more on the thermal aspect of the structure of the turbulent flow through a heterogeneous fractal-like forest canopy. 

Turbulent jets generated by a uniform flow passing through a fractal structure at inflow have received attention in a variety of settings. For instance, Cafiero et al.\cite{Cafiero2014} studied impinging jets and found that a fractal turbulence promoter can provide a significant heat transfer enhancement for relatively small nozzle-to-plate separation compared to that produced by a regular grid in the same power input conditions (see also Cafiero et al. \cite{Cafiero2015,Cafiero2016}). 
Other studies showed that multi-scale fractal geometries affect the large-scale coherent structures of an axisymmetric turbulent jet (see Breda and Buxton \cite{Breda_and_Buxton_1,Breda_and_Buxton_2}).
Moreover, it has been shown that the highly turbulent flow generated by the fractal structures reduces the impact of re-circulation flow (Nedi$\acute{c}$ et al. \cite{Nedic}, Rodr\'{i}guez-L\'{o}pez et al. \cite{Lopez}). Nedi$\acute{c}$ et al. \cite{Nedic2012} studied the aeroacoustic performance of scaled fractal spoilers. The results of these studies showed that a fractal square spoiler can reduce the noise (up to 4dB) whilst not affecting the lift and drag characteristics. They also noticed a highly intense bleed flow as well as an increase in the levels of turbulence intensity, 2D TKE (defined in\cite{Nedic2012} as $0.5*(u^{'2}+w^{'2})$), and velocity compared to a regular grid spoiler. Breda and Buxton \cite{Breda_and_Buxton_1,Breda_and_Buxton_2} obtained similar results for fractal jets. They showed that the fractal geometry significantly changes the near-field structure of the jet, breaking up the large-scale coherent structures, responsible for the low-frequency noise, and thereby affecting the entrainment rate of the background fluid into the jet stream. 

Other interesting applications of fractal geometries have been considered in the area of combustion mixing. Geipel et al. \cite{Geipel} conducted a parametric study targeting turbulence generation in a jet geometry combined with fractal grids, and compared the results to those obtained using conventional plates; they showed that fractal grids can increase turbulence levels by more than 100\% (see also Goh et al. \cite{Goh} for a followup study). Through a series of experiments, Soulopoulos et al. \cite{Soulopoulos} showed that by using a fractal grid instead of a regular square mesh grid in a burner configuration, the turbulent burning velocity of the flame stabilizes, and is higher than the burning velocity corresponding to the square mesh grid. In a later study performed by Verbeek et al. \cite{Verbeek}, it was shown that replacing a classical hexagonal grid by
a multiscale fractal grid yields an increased turbulent flame speed (a follow-up study was conducted by Thij et al. \cite{Thij}).


Thus far, studies in the area of interaction of flows with fractal geometries mainly focused on the incompressible regime. The present study therefore aims at investigating the effect of fractal geometries on compressible flows via LES. Our main focus is to validate against the experimental work of Nedi$\acute{c}$ \cite{Nedic1} on wake generation downstream of flow through fractal plates and extend it to higher Mach numbers. The numerical algorithm is based on high-order low-dissipation spatial and temporal schemes in an implicit LES framework. The square and fractal structures are modeled using an immersed boundary method, with adequate grid resolution in proximity to the small-scale fractal petals as shown in figure \ref{f1}. We validate our numerical results via extensive comparisons to experimental measurements at a low Mach number (Nedic \cite{Nedic1}). Here we consider profiles of the mean flow, Reynolds stresses in the transverse direction, and TKE distribution along the centerline. The subsequent numerical investigation focuses on a wide range of inflow Mach numbers covering both the subsonic and supersonic regimes.


The rest of the paper is organized as follows: Section II presents the governing equations and describes the numerical framework employed with a brief introduction of the adopted terminology. In Sec. III, we compare our numerical results to Nedic's experimental measurements to validate the numerical algorithm at a Mach number of 0.08. In Sec. IV, we report our numerical results for various diagnostic turbulence quantities such as two-point velocity correlations, energy and pressure spectra for the multiscale fractal-generated wakes at Mach numbers ranging from $0.2$ to $2.5$. Sec. V includes a summary of the results and several conclusions.

\section{Large eddy simulation framework}\label{}

\subsection{Scalings}
  
  The governing equations involve a generalized curvilinear coordinate transformation, which is written in the three-dimensional form as
  $\xi = \xi \left(x,y,z \right),
  \eta = \eta \left(x,y,z \right),
  \zeta = \zeta \left(x,y,z \right)$,
  where $\xi$, $\eta$ and $\zeta$ are the spatial coordinates in the computational space corresponding to the streamwise, wall-normal and spanwise directions, and $x$, $y$ and $z$ are the spatial coordinates in physical space. This transformation allows for a seamless mapping of the solution from the computational to the physical space and vice-versa. All dimensional spatial coordinates are normalized by the reference length $D$ associated with the fractal geometry (the square root of the plate area),
  \begin{eqnarray}\label{NS}
  (x,y,z) = \frac{(x^*,y^*,z^*)}{D},
  \end{eqnarray}
  the velocity is scaled by the freestream velocity magnitude $V_{\infty}^*$, 
  \begin{eqnarray}\label{NS}
  (u,v,w) = \frac{(u^*,v^*,w^*)}{V_{\infty}^*},
  \end{eqnarray}
  the pressure by the dynamic pressure at infinity, $\rho_{\infty}^* V_{\infty}^{*2}$, and temperature by the freestream temperature, $T_{\infty}^*$. Reynolds number based on $D$ and the freestream velocity, Mach number, and Prandtl number are defined as
  \begin{eqnarray}\label{NS}
  R_e = \frac{\rho_{\infty}^* V_{\infty}^* D}{\mu_{\infty}^*}, \hspace{5mm}
  Ma = \frac{V_{\infty}^*}{a_{\infty}^*}, \hspace{5mm}
  Pr = \frac{\mu_{\infty}^* C_p}{k_{\infty}^*}
  \end{eqnarray}
  where $\mu_{\infty}^*$, $a_{\infty}^*$ and $k_{\infty}^*$ are freestream dynamic viscosity, speed of sound and thermal conductivity, respectively, and $C_p$ is the specific heat at constant pressure. All simulations are performed for air as an ideal gas.

\subsection{Governing equations}\label{}

 In conservative form, the filtered Navier-Stokes equations are written as
  \begin{eqnarray}\label{NS}
  \mathbf{Q}_t
  + \mathbf{F} _{\xi}
  + \mathbf{G}_{\eta}
  + \mathbf{H}_{\zeta}
  = \mathbf{S}.
  \end{eqnarray}
  where subscript denote derivatives, the vector of conservative variables is given by
  \begin{equation}
  \mathbf{Q} = \frac{1}{J} \{ 
  \begin{array}{rrrrrr}
  \rho,   \hspace{4mm}
  \rho u_i,   \hspace{4mm}
  E
  \end{array}
  \}^{T}, i = 1,2,3
  \end{equation}
  where $J$ is the Jacobian, $\rho$ is the non-dimensional density of the fluid, $u_i = (u, v, w)$ is the non-dimensional velocity vector in physical space, and $E$ is the total energy. The flux vectors, $\mathbf{F}$, $\mathbf{G}$ and $\mathbf{H}$, are given by
  
  \begin{align}\label{}
    & \hspace{3.5mm}\mathbf{F} = \frac{1}{J} \left\{ \rho U,\hspace{0.5mm} \rho u_iU + \xi_{x_i} (p + \tau_{i1}),\hspace{0.5mm} E U + p \tilde{U} +  \xi_{x_i} \Theta_i\right\}^{T}, \\
    &\quad \mathbf{G} = \frac{1}{J} \left\{\rho V,\hspace{0.5mm}\rho u_iV + \eta_{x_i} (p + \tau_{i2}), \hspace{0.5mm}E V + p \tilde{V}+  \eta_{x_i} \Theta_i\right\}^{T},  \\ 
    &\quad \mathbf{H} = \frac{1}{J} \left\{\rho W,\hspace{0.5mm}\rho u_iW + \zeta_{x_i} (p + \tau_{i3}),\hspace{0.5mm}E W + p \tilde{W}+  \zeta_{x_i} \Theta_i\right\}^{T}
  \end{align}
  where the contravariant velocity components are given by
  \begin{eqnarray}\label{con}
  U = \xi_{x_i} u_i ,   \hspace{4mm}
  V = \eta_{x_i} u_i,    \hspace{4mm}
  W = \zeta_{x_i} u _i
  \end{eqnarray}
  with the Einstein summation convention applied over $i = 1,2,3$, the shear stress tensor and the heat flux are respectively given as
  
  \begin{equation}
  \tau_{ij} = \frac{\mu}{Re} \left[
  \left(
  \frac{\partial \xi_k}{\partial x_j}  \frac{\partial u_i}{\partial \xi_k}  +
  \frac{\partial \xi_k}{\partial x_i}  \frac{\partial u_j}{\partial \xi_k}
  \right)
  - \frac{2}{3} \delta_{ij} \frac{\partial \xi_l}{\partial x_k}  \frac{\partial u_k}{\partial \xi_l}
  \right]
  \end{equation}
  
  \begin{equation}
  \Theta_{i} = 
  u_j \tau_{ij} + \frac{\mu}{(\gamma-1)M_{\infty}^2 Re Pr}
  \frac{\partial \xi_l}{\partial x_i}  \frac{\partial T}{\partial \xi_l}
  \end{equation} 
  $\mathbf{S}$ is the source vector term, and $\mu$ is the dynamic viscosity.
   
  The pressure $p$, the temperature $T$  and the density of the fluid are combined in the equation of state, $p = \rho T / \gamma M_{\infty}^2$. The Jacobian of the curvilinear transformation from the physical space to computational space is denoted by $J$. The derivatives $\xi_x$, $\xi_y$, $\xi_z$, $\eta_x$, $\eta_y$, $\eta_z$, $\zeta_x$, $\zeta_y$, and $\zeta_z$ represent grid metrics. The dynamic viscosity is linked to the temperature using Sutherland's equation in dimensionless form,
  \begin{eqnarray}
  \mu = T^{3/2} \frac{1 + C_1/T_{\infty}}{T+C_1/T_{\infty}},
  \end{eqnarray}
  while the thermal conductivity $k$ is obtained from the Prandtl number, where for air at sea level, $C_1 = 110.4 K$. There are no subgrid scale terms in equation (\ref{NS}) since an implicit large eddy simulation framework is considered here.

\subsection{Numerical algorithm}\label{}

The compressible Navier-Stokes equations are solved in the framework of implicit large eddy simulations, where numerical filtering is applied to account for the missing sub-grid scale energy. The numerical algorithm uses high-order finite difference approximations for the spatial derivatives and explicit time marching. The time integration is performed using a third order TVD Runge-Kutta method (Shu and Osher \cite{Shu}). 
  
  The spatial derivatives are discretized using dispersion-relation-preserving schemes of Tam and Webb \cite{Tam} or a high-resolution 9-point dispersion-relation-preserving optimized scheme of Bogey et al. \cite{Bogey}.  
To damp out the unwanted high wavenumber waves from the solution, high-order spatial filters, as developed by Kennedy and Carpenter \cite{Kennedy}, are applied. A no slip boundary condition for velocity and adiabatic condition for temperature are imposed at the solid surface through the immersed boundary method. Sponge layers are imposed in proximity to the far-field boundaries, regions that are outside the flow domain of interest (this is combined with grid stretching to damp the unwanted waves). These sponge layers are designed to damp out the waves of all wavenumbers exiting the domain or reflecting back from the boundaries.
  
Shock capturing techniques are implemented to avoid unwanted oscillations that may propagate from potential discontinuities that can develop in supersonic flows. We apply a shock capturing methodology that was proven to work efficiently for high-order, nonlinear computations (Bogey et al. \cite{Bogey2}). Since in the present work high-order, central-difference schemes are used to achieve increased resolution of the propagating disturbances, a straightforward approach is a model which introduces sufficient numerical viscosity in the area of the discontinuities, and negligible artificial viscosity in the rest of the domain. A specific shock-capturing technique, suitable for simulations involving central differences in space is applied, based on the general explicit filtering framework. The technique introduces selective filtering at each grid vertex to minimize numerical oscillations, and shock-capturing in the areas where discontinuities are present (more details can be found in Bogey et al. \cite{Bogey2}).
  
The fractal structures will be modeled using an immersed boundary method (IBM) relying on a penalization approach. The first recorded IBM was employed by Peskin \cite{Peskin,Peskin_1} to simulate the interaction of the blood flow with a heart valve, modeled as a non-conforming elastic solid. Many other researchers developed extensions of this method to handle solid boundaries in the rigid limit. Since these approaches suffer from stiffness problems in the limit of rigid solids as a result of the high gradients of the forcing term, high spatial resolution is needed to obtain an accurate representation of the interface. To overcome these issues, other authors proposed an alternative class of IBM in the form of penalization methods (Angot et al. \cite{Angot}, Vincent et al. \cite{Vincent_1}).  This approach proved high efficiency with complex geometries such as those encountered in fish-like swimming applications (\cite{Bergmann}, \cite{Gazzola}, and \cite{Ghaffari}). Other applications of this method include areas dealing with compressible flows (\cite{Boiron} and \cite{Piquet}), combined with level-set methods for interface tracking problems (\cite{Bergmann}, \cite{Chantalat} and \cite{Gazzola}), fluid-structure interactions problems as in \cite{Huang} or \cite{Engels}, and highly turbulent flows (\cite{Bergmann} and \cite{Vincent_2}).
  
The volume penalization IBM renders easier simulating flow past complex geometries, such as multi-scale fractal structures, using a non-body conformal Cartesian grid instead of employing structured or unstructured grids that conform to the body. In this approach, the Cartesian volume grid is generated regardless of the immersed solid boundary surface grid, which only cuts through the Cartesian grid. The immersed boundary method in the framework of flows interactions with fractal structures was already utilized by Suzuki et al. \cite{Suzuki_1, Suzuki_3}, Laizet and Vassilicos \cite{Laizet2011}, Chester et al. \cite{Chester}, Graham et al. \cite{Graham}. Dairay et al. \cite{Dairay} performed a combined direct numerical simulation and hot-wire anemometry investigation
of turbulent wakes generated by square and fractal plates similar to the ones that are considered in the current study.
  

\subsection{Numerical simulation settings}\label{}

\begin{figure}
 \begin{center}
   \includegraphics[width=3.5cm]{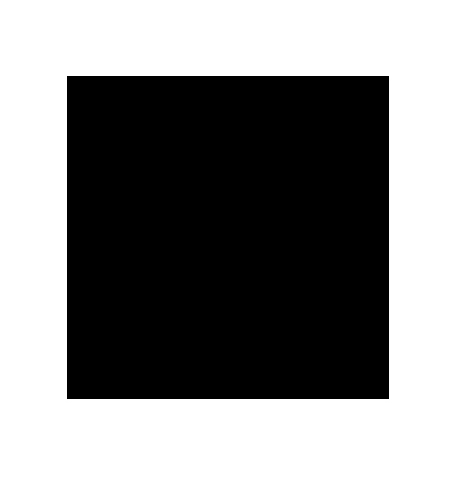} 
   \hspace{8mm}
   \includegraphics[width=3.7cm]{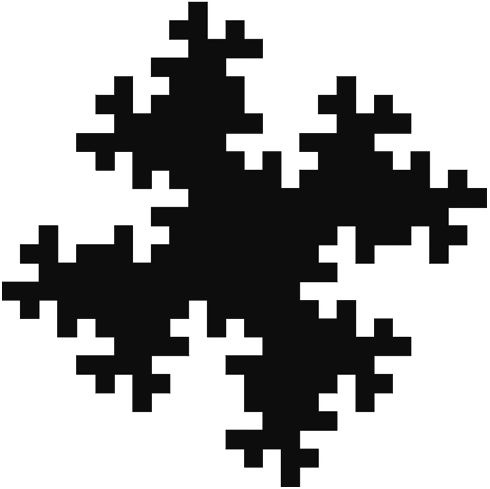} \\
   a) \hspace{40mm} b) 
 \end{center}
  \caption{\label{} Flat plates used in the analysis: a) square; b) fractal $Df=1.5(2)$.}
  \label{f1}
\end{figure}

In this study, we consider two thin plates that are normal to the flow direction; the first has a square shape, and the second represents a multi-scale fractal (see figure \ref{f1}). Both plates have the same frontal area and interact with subsonic and supersonic flows at Mach numbers 0.2, 0.7, 1.4, and 2.5. Both plates have the same thickness of 0.06 fraction from the side of the square to prevent the development of boundary layers on the surface. The characteristic length of the plates $L=\sqrt{A}$, which is equal to the side of the square. 

The structured grid consists of $160$ blocks with a total of $2.7 \times 10^{7}$ grid points with an increased resolution in the vicinity of the plates, and stretching towards the far-field boundaries. As mentioned previously, the effect of the plates on the flow is modeled via an immersed boundary method, with the grid resolution around the smallest scale of the fractal geometry being approximately 5-6 points per the smallest side. Hereafter we will refer to the multi-scale fractal plate using the terminology introduced by Nedi$\acute{c}$ \cite{Nedic1}, that is $Df = 1.5(2)$, where the first number represents the fractional dimension of the fractal geometry (given as $D_f=log(d^n)/log(\lambda/l_n)$, where $l_n=\lambda/r^n$, $r=4(cos \alpha - 1)+d$, $r$ is the ratio of the successive lengths, $n$ is the iteration count, $\lambda$ is the base straight line length, and $d$ is the number of segments) as opposed to the finite dimension (2D) of the square, while the second number represents the number of iterations used to obtain the fractal. 

\begin{figure}
 \begin{center}
   \includegraphics[width=6.0cm]{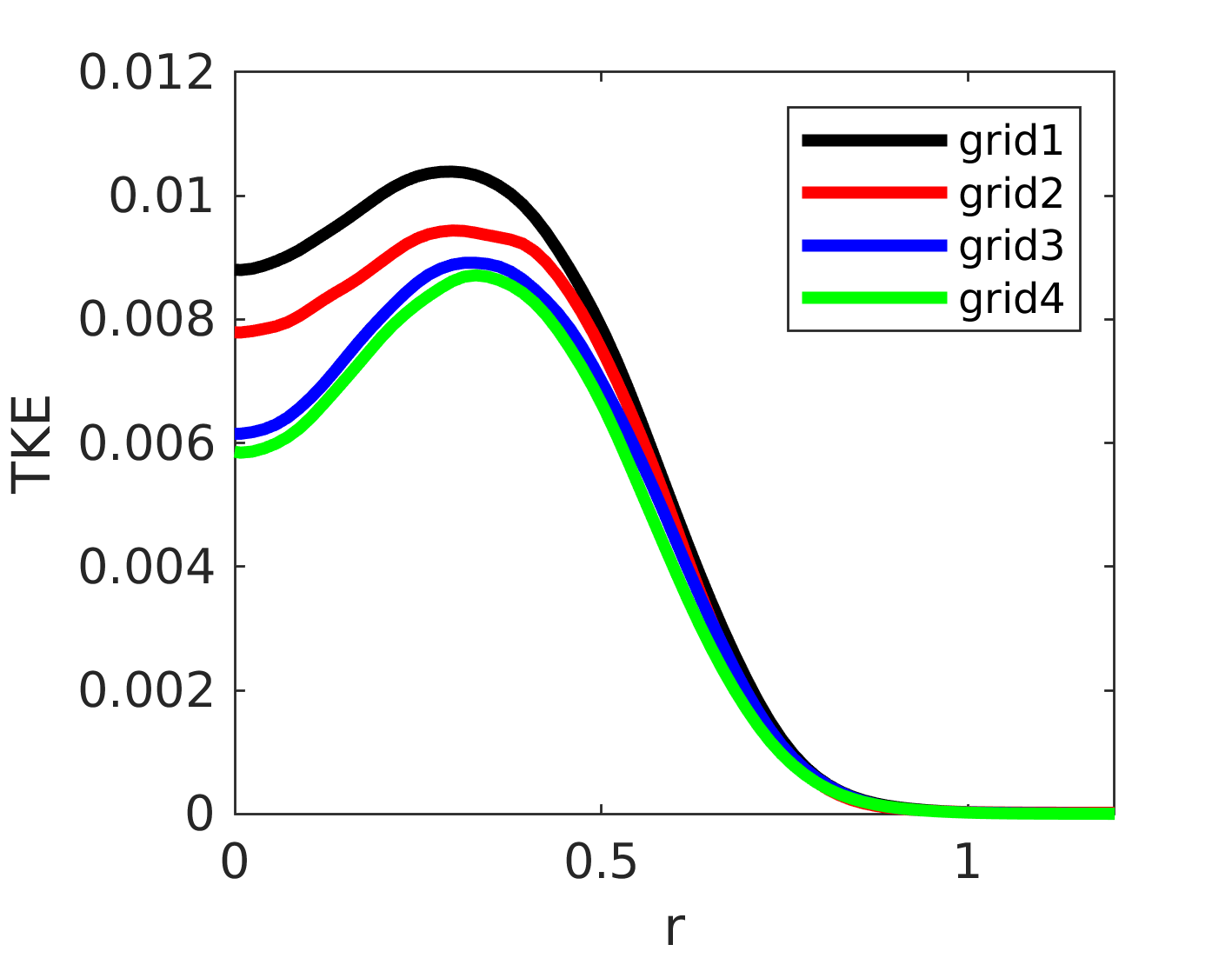}
 \end{center}
  \caption{\label{} Comparison of TKE profiles in $x=5$ for four grid resolutions.}
  \label{grid}
\end{figure}

We conducted a grid sensitivity analysis to ensure that the LES accurately resolves the flow in the wake. We ran simulations of four grid resolutions (9, 16, 22, and 27 million points, respectively) for the fractal plate with the flow Mach number set to $1.4$. Figure \ref{grid} illustrates a close matching in the cross-flow distributions of the TKE for grid 3 and 4, which indicates the adequacy of the chosen grid resolution (grid 4) to resolve and adequately capture the details of the near flow-structures.

\section{Comparison to experimental measurements in the incompressible regime}\label{}

We validate our numerical algorithm through comparisons to wind tunnel experimental measurements in the incompressible regime, conducted by Nedi$\acute{c}$ \cite{Nedic1}. The wind tunnel had a working cross section of $0.91$m $\times$ $0.91$m and a length of $4.88$m. The experiments were conducted at a freestream velocity of $10$ m/s, with a background turbulence level of $0.05\%$. In our simulations, the Mach number was set to $0.08$ ($27.2$ m/s) due to some limitations associated with the compressible solver; however, we maintained the same Reynolds number, $R_e = 41,000$, as in the experiments, which corresponds to a square of side $2.4$ cm.

\begin{figure}
 \begin{center}
   \includegraphics[width=5cm]{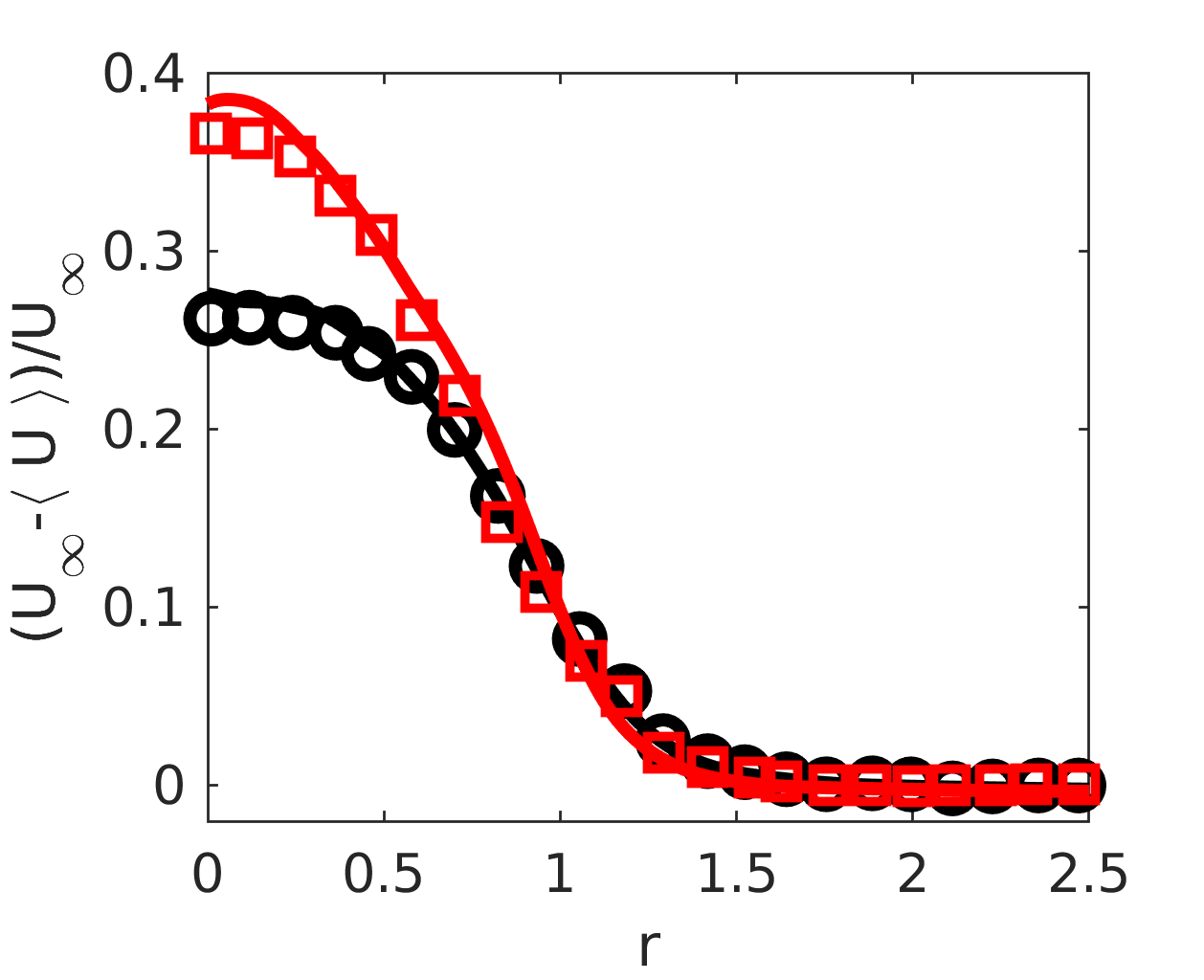}
   \includegraphics[width=5cm]{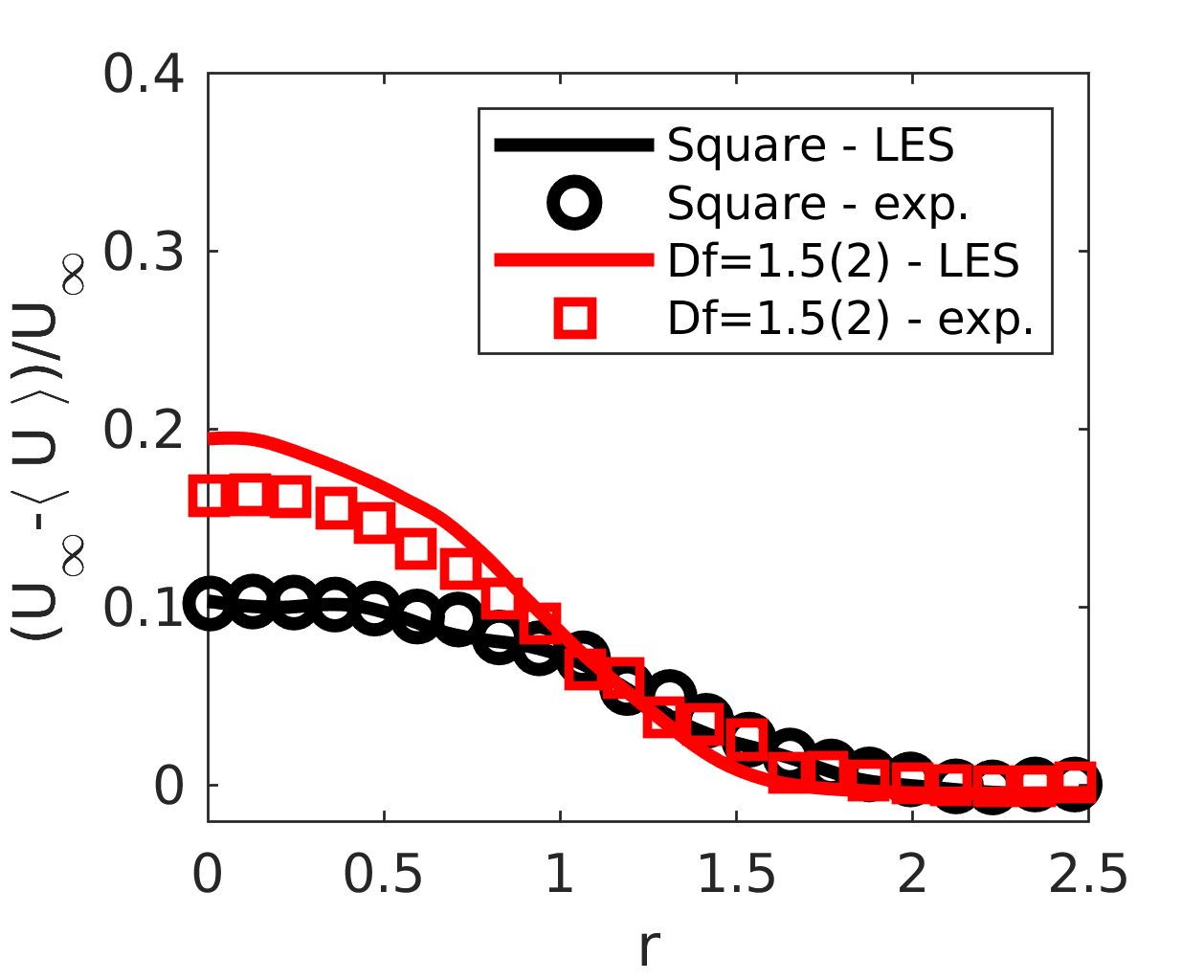} \\
   \hspace{0.35cm} (a) \hspace{4.5cm} (b)\\
 \end{center}
  \caption{\label{} Mean streamwise velocity compared to experimental measurements of Nedi$\acute{c}$ \cite{Nedic1}: a) x=5; b) x = 10.}
  \label{f6}
\end{figure}

We compare the mean streamwise velocity deficit $(U_{\infty} - \langle U \rangle)/U_{\infty}$ distribution in the crossflow direction to experimental measurements in figure \ref{f6} at two streamwise locations, $x=5$ and $x=10$. The agreement between the numerical and experimental results at $x=5$ is very good. At $x=10$, the results corresponding to the square plate match, while the result from the numerical simulation for the fractal plate slightly over predicts the experimental data. The size of the wake, quantified by the radial location where the mean velocity becomes zero, is captured accurately by the numerical simulations.

\begin{figure}
 \begin{center}
   \includegraphics[width=5cm]{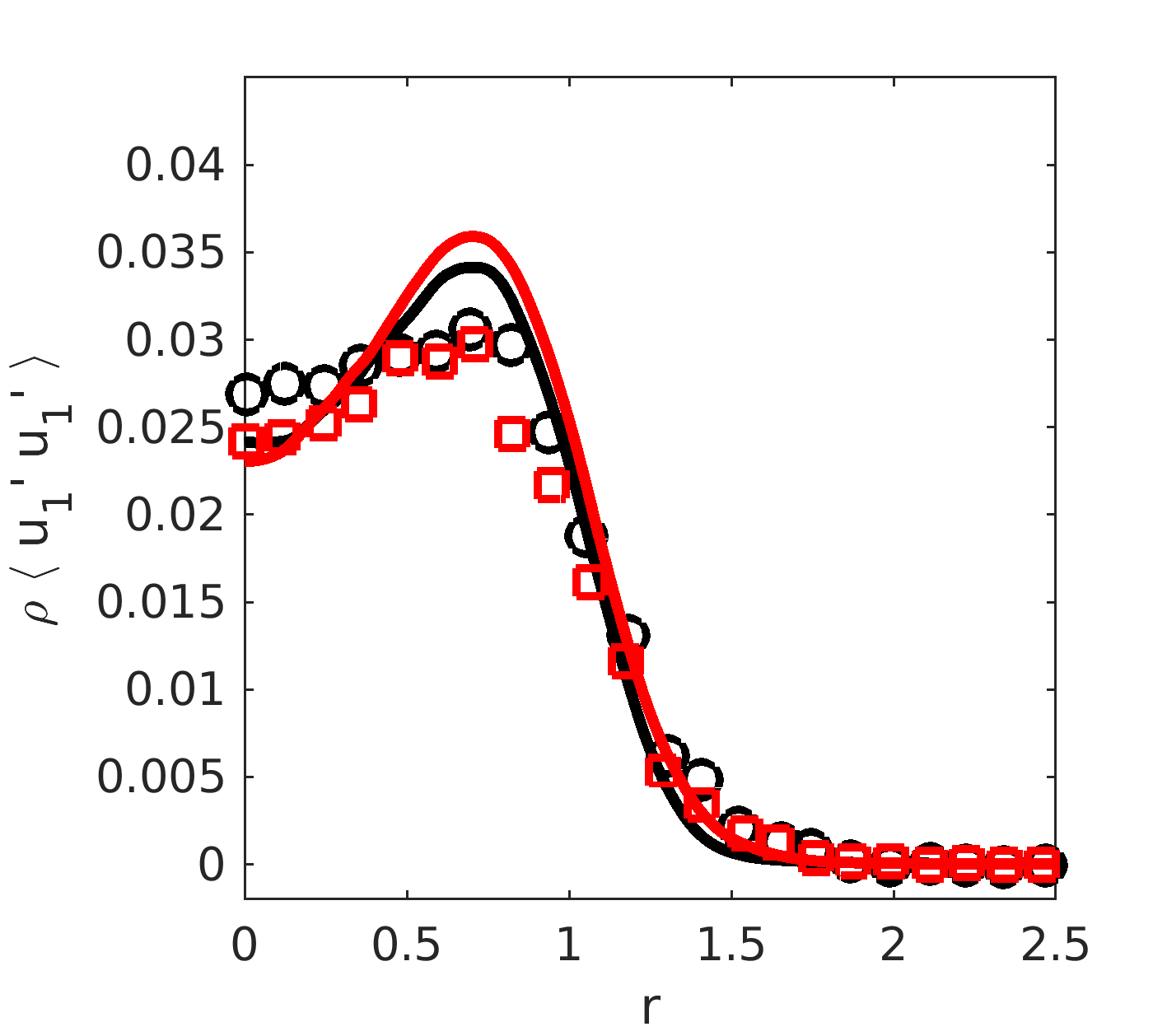}
   \includegraphics[width=5cm]{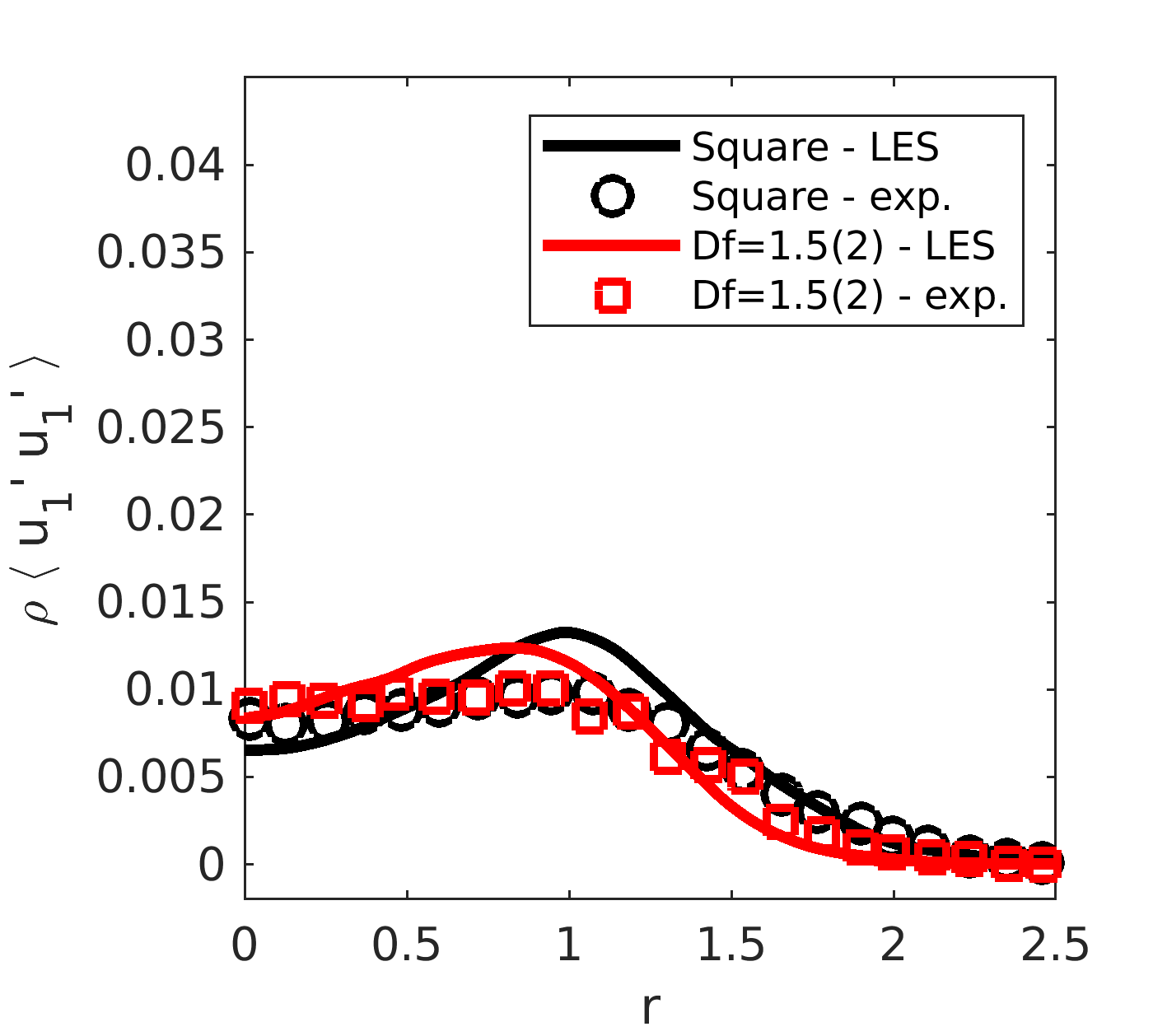} \\
   \hspace{0.35cm} (a) \hspace{4.5cm} (b)\\
 \end{center}
  \caption{\label{} Reynolds stress tensor component $\rho\langle u_1'u_1' \rangle$ compared to experimental measurements of Nedi$\acute{c}$ \cite{Nedic1}: a) x=5; b) x = 10.}
  \label{f7}
\end{figure}

\begin{figure}
 \begin{center}
   \includegraphics[width=5cm]{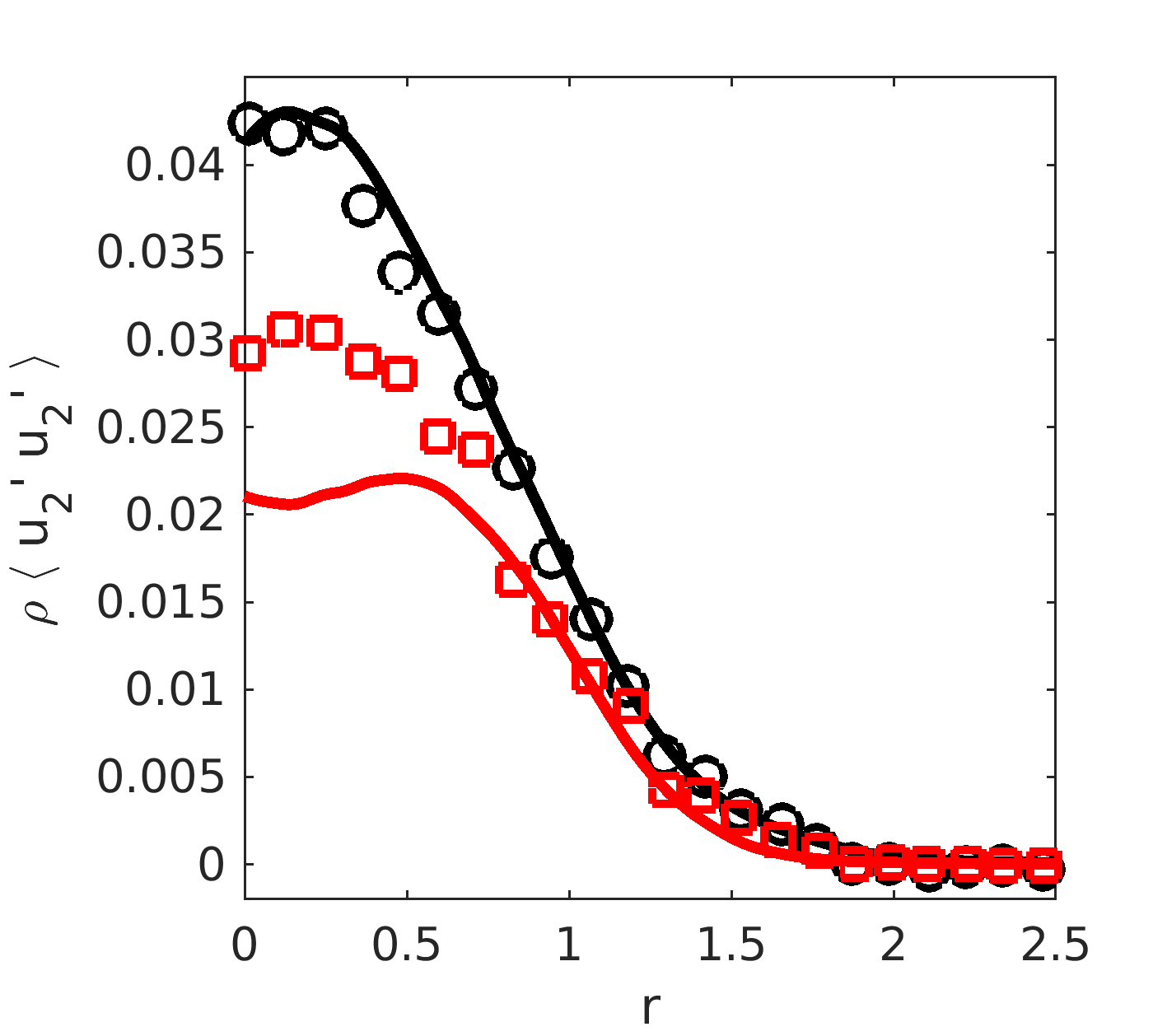}
   \includegraphics[width=5cm]{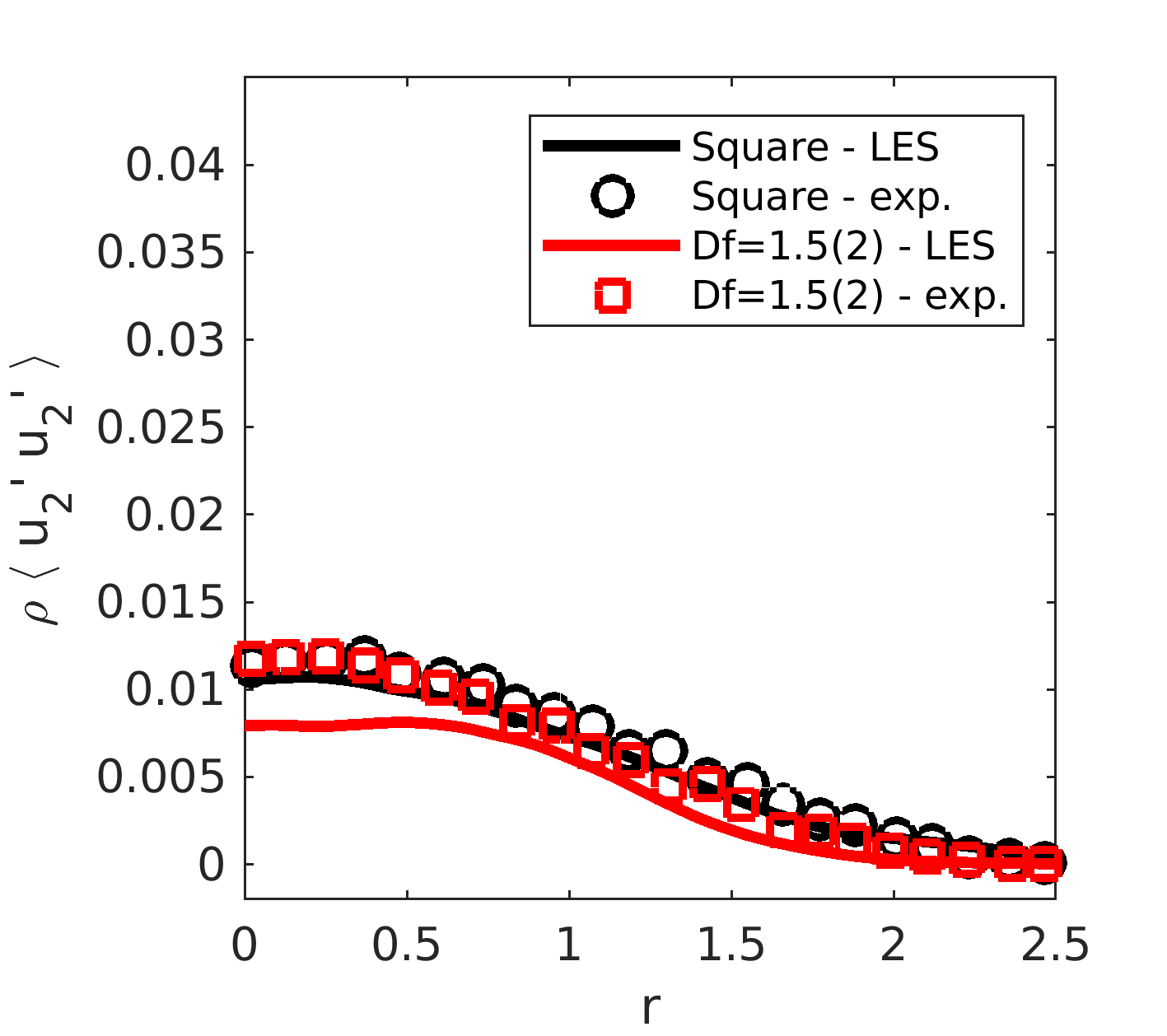} \\
   \hspace{0.35cm} (a) \hspace{4.5cm} (b)\\
 \end{center}
  \caption{\label{} Reynolds stress tensor component $\rho\langle u_2'u_2' \rangle$ compared to experimental measurements of Nedi$\acute{c}$ \cite{Nedic1}: a) x=5; b) x = 10.}
  \label{f8}
\end{figure}

We also compare the Reynolds stress components $\rho\langle u_1'u_1' \rangle$ and $\rho\langle u_2'u_2' \rangle$ to experimental data along the radial direction in figures \ref{f7} and \ref{f8}, respectively. The overall levels of $\rho\langle u_1'u_1' \rangle$ for the two geometries in figure \ref{f7} are close to each other for both streamwise locations, $x=5$ and $x=10$, although there is a slight over prediction in the shear layer (this was also observed in the DNS results of Dairay et al. \cite{Dairay}). The $\rho\langle u_2'u_2' \rangle$ component in figure \ref{f8} shows a good agreement for the square plate, and an under prediction for the fractal plate, at both streamwise locations; this behavior might either be a result of the grid resolution, or from not imposing adequate freestream disturbances on the incoming flow, as in the experimental measurements. The decay to zero of both predicted Reynolds stress components as $r \rightarrow \infty$ displays good agreement with the experimental data.

\begin{figure}
 \begin{center}
   \includegraphics[width=8cm]{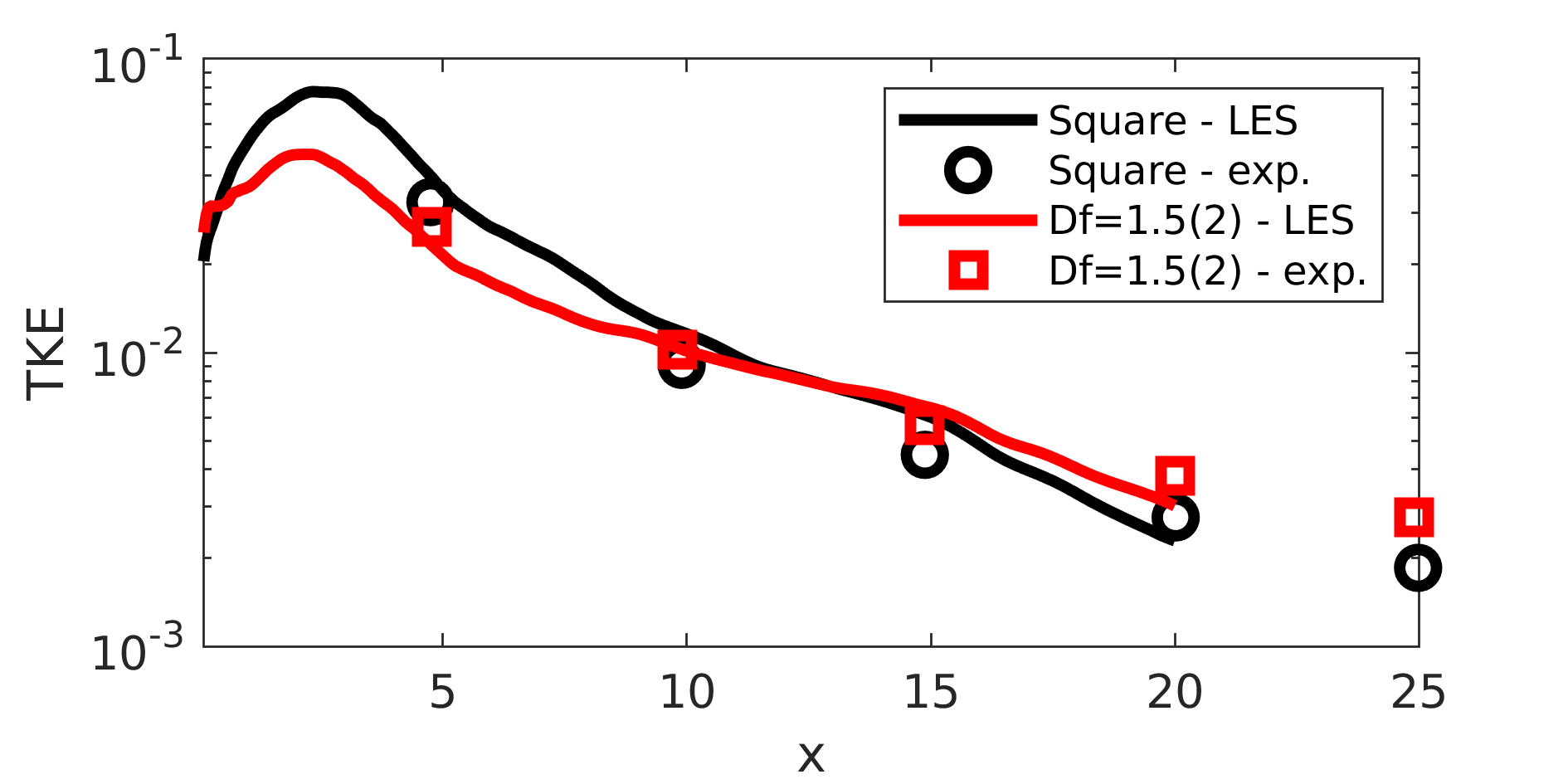}
 \end{center}
  \caption{\label{} TKE along the centerline compared to experimental measurements of Nedi$\acute{c}$ \cite{Nedic1}.}
  \label{f9}
\end{figure}

In figure \ref{f9}, we compare our centerline TKE distribution to experimental measurements for $x$ between $0$ and $20$, in logarithmic scale on the vertical direction (experimental data was obtained from $x=5$). The agreement between the two sets of data was found to be good; the numerical results also captured the `switching' point, where the TKE corresponding to the fractal plate becomes larger than the TKE of the square plate (around $x=12$).

\section{Numerical Results for Flows at Higher Mach Number}\label{}

In this section, we analyze the effect of the plates considered in figure \ref{f1} on compressible subsonic and supersonic freestream flows. The results consist of iso-surfaces of {\color{black} vorticity magnitude}, contour plots of pressure, mean flow, Reynolds stresses, TKE in the crossflow or streamwise direction, longitudinal and transverse correlations, as well as velocity and pressure spectra.

\subsection{Iso-surfaces and contour plots}\label{}

\begin{figure}
 \begin{center}
   \includegraphics[width=10.0cm]{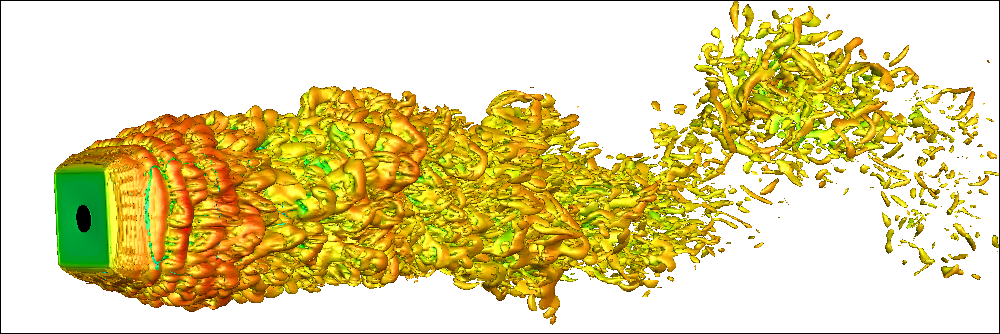} \\
   \includegraphics[width=10.0cm]{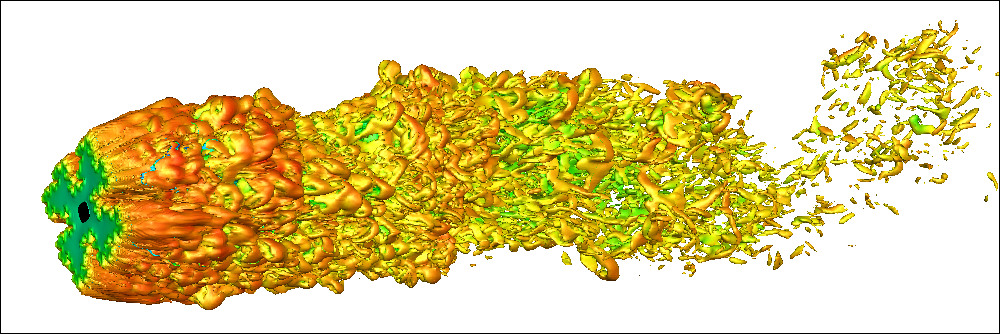} 
 \end{center}
  \caption{\label{} Iso-surfaces of {\color{black} vorticity magnitude} colored by the streamwise velocity for Mach number 0.2: square (top) and $Df=1.5(2)$ (bottom) plates.}
  \label{f2}
\end{figure}

\begin{figure}
 \begin{center}
   \includegraphics[width=10.0cm]{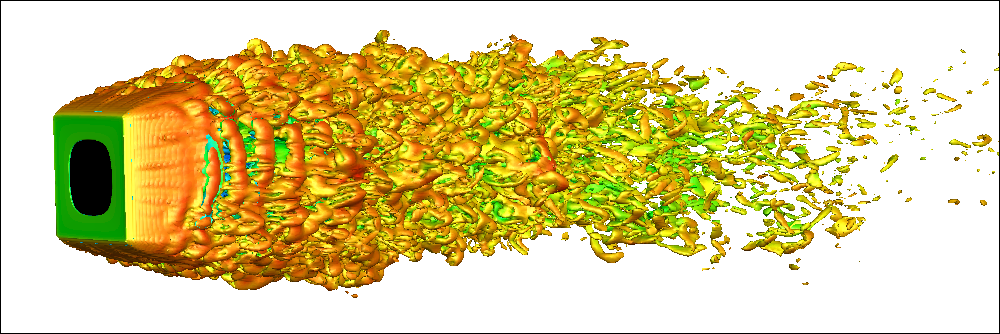} \\
   \includegraphics[width=10.0cm]{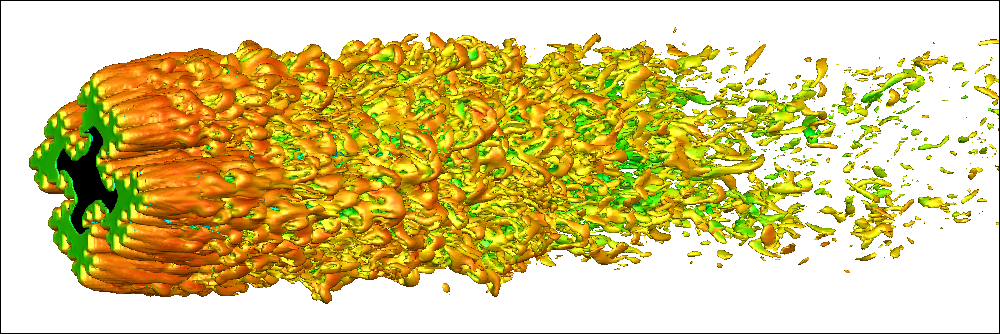} 
 \end{center}
  \caption{\label{} Iso-surfaces of {\color{black} vorticity magnitude} colored by the streamwise velocity for Mach number 0.7: square (top) and $Df=1.5(2)$ (bottom) plates.}
  \label{f3}
\end{figure}

\begin{figure}
 \begin{center}
   \includegraphics[width=10.0cm]{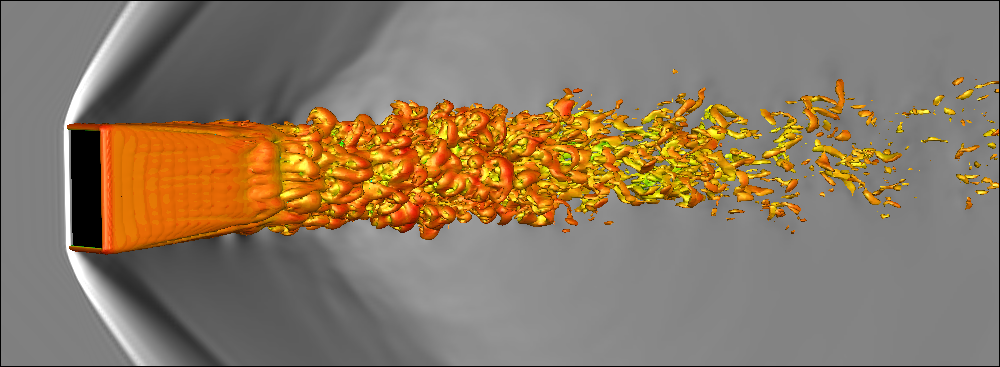} \\
   \includegraphics[width=10.0cm]{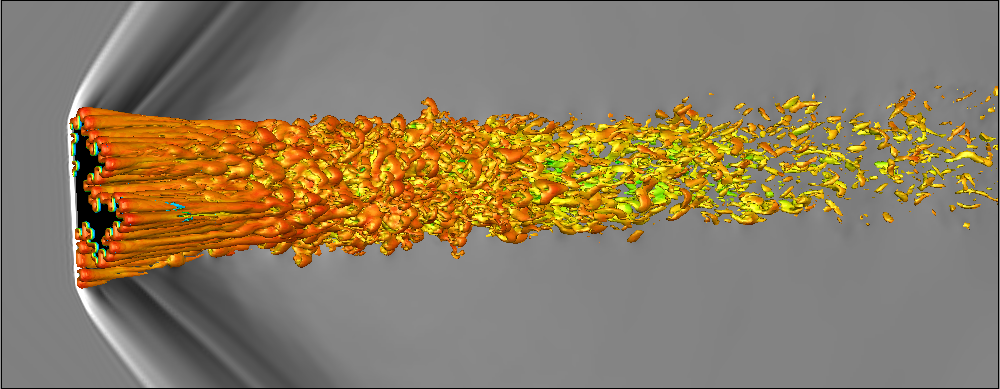} 
 \end{center}
  \caption{\label{} Iso-surfaces of {\color{black} vorticity magnitude} colored by the streamwise velocity and contours of pressure (in gray) for Mach number 1.4: square (top) and $Df=1.5(2)$ (bottom) plates.}
  \label{f4}
\end{figure}

\begin{figure}
 \begin{center}
   \includegraphics[width=10.0cm]{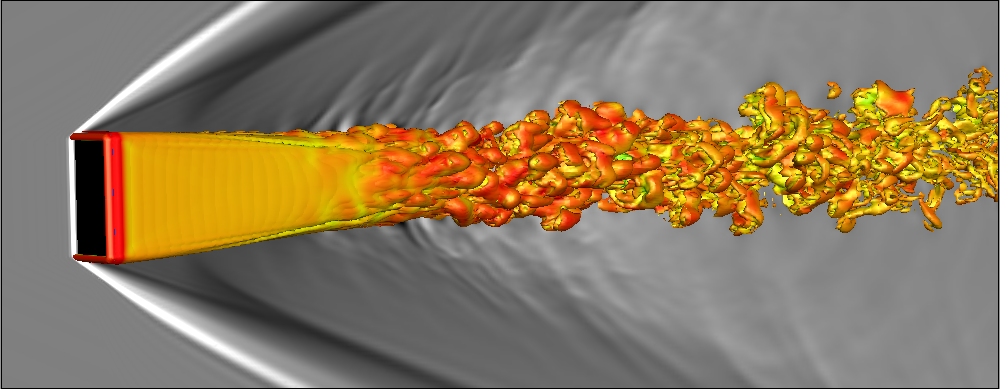} \\
   \includegraphics[width=10.0cm]{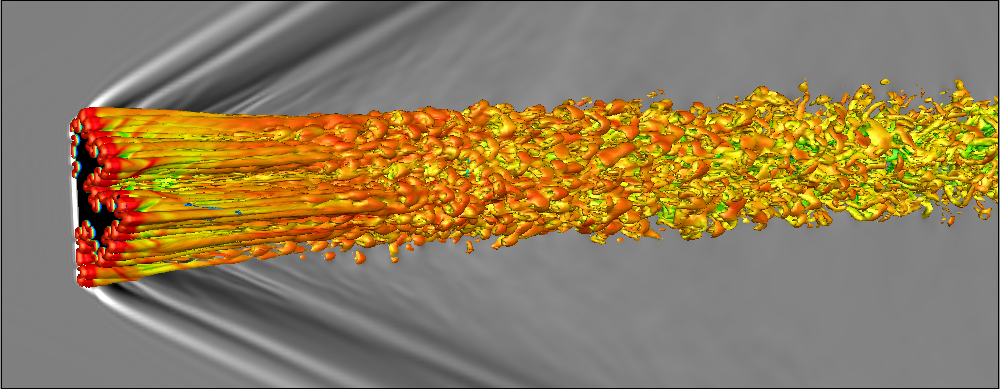} 
 \end{center}
  \caption{\label{} Iso-surfaces of {\color{black} vorticity magnitude} colored by the streamwise velocity and contours of pressure (in gray) for Mach number 2.5: square (top) and $Df=1.5(2)$ (bottom) plates.}
  \label{f5}
\end{figure}

In figures \ref{f2}-\ref{f5}, we show iso-surfaces of {\color{black} vorticity magnitude} colored by the streamwise velocity component. The {\color{black} vorticity magnitude} has been normalized and plotted for a non-dimensional value $1$ for all Mach numbers. {\color{black} The wakes develop in the downstream along the positive x-axis, which is perpendicular to the plate, with y-axis pointing upward; the plate is located at $x=0$.} Figure \ref{f2}, corresponding to the lowest Mach number of $0.2$, provides a qualitative representation of the wake evolving downstream for both plates. One important observation, which is consistent with previous findings, is that in the case of the square plate there are organized flow structures in proximity to the plate, while for the fractal plate these structures are broken up into smaller fluctuations. Wake meandering is noticed for both cases, although it seems less intense in the fractal plate case, which is presumably the result of increased mixing which suppresses the formation of large scale structures downstream (wake meandering was also observed in the DNS results of Dairay et al. \cite{Dairay} through iso-surfaces of vorticity magnitude, although the results for the square plate were not reported).

In figure \ref{f3}, we show iso-surfaces of {\color{black} vorticity magnitude} for the subsonic compressible regime of Mach number $0.7$. The same observation with respect to breaking up the flow structures by the fractal shapes holds here, except there is no wake meandering (this phenomenon is characteristic to incompressible, low-speed flows). Azimuthal instabilities developing in close proximity to the square plate can be observed for both Mach numbers of $0.2$ in figure \ref{f2} and $0.7$ in figure \ref{f3}, both of which transition into turbulence further downstream; these types of instabilities are not seen in the fractal plate case because the small scale fractal structures break down these instabilities very efficiently.

The next two figures, \ref{f4} and \ref{f5}, correspond to supersonic flows of $M=1.4$ and $M=2.5$, respectively, interacting with the plates, and apart from iso-surfaces of {\color{black} vorticity magnitude} they also illustrate contour plots of pressure. The contour plots of pressure reveal the existence of shock waves, and provide a qualitative representation of the acoustic {\color{black} pressure} field generated by these flows (the noise emitted from these high Mach number flows is expected to be high). Downstream of the square plate, a long region of {\color{black} low turbulence intensity (as will be revealed in subsequent figures)} develops for both {\color{black} supersonic} Mach numbers. {\color{black} The instabilities characterizing this region have smaller amplitudes compared to the those of the subsonic cases, therefore, triggering the transition into turbulence much further downstream}. Looking at the difference between the two supersonic cases in terms of the length of their laminar regions, there is reason to believe that the length of this transition region increases with the Mach number. The wake generated by the interaction with the fractal plate displays features of a fully turbulent flow along the entire length of the wake (the transition from laminar to turbulent flow takes place in the vicinity of the plate). 

%

%
{\color{black} Pressure contours plotted in gray in the external region of the wakes in figures \ref{f4} and \ref{f5} are indicative of the} level of {\color{black} aerodynamic sound} that is generated by the transition region in the wake of the square plate. {\color{black} These pressure waves} are more intense for the higher Mach number of $2.5$ in figure \ref{f5}. The {\color{black} pressure waves} generated by the flow interaction with the fractal plates, however, {\color{black} are} less intense. A slight weakening of the shock waves can be also noticed when comparing the square to the fractal plate. 
\subsection{Mean flow and Reynolds stresses}\label{}

\begin{figure}
 \begin{center}
   \includegraphics[width=5cm]{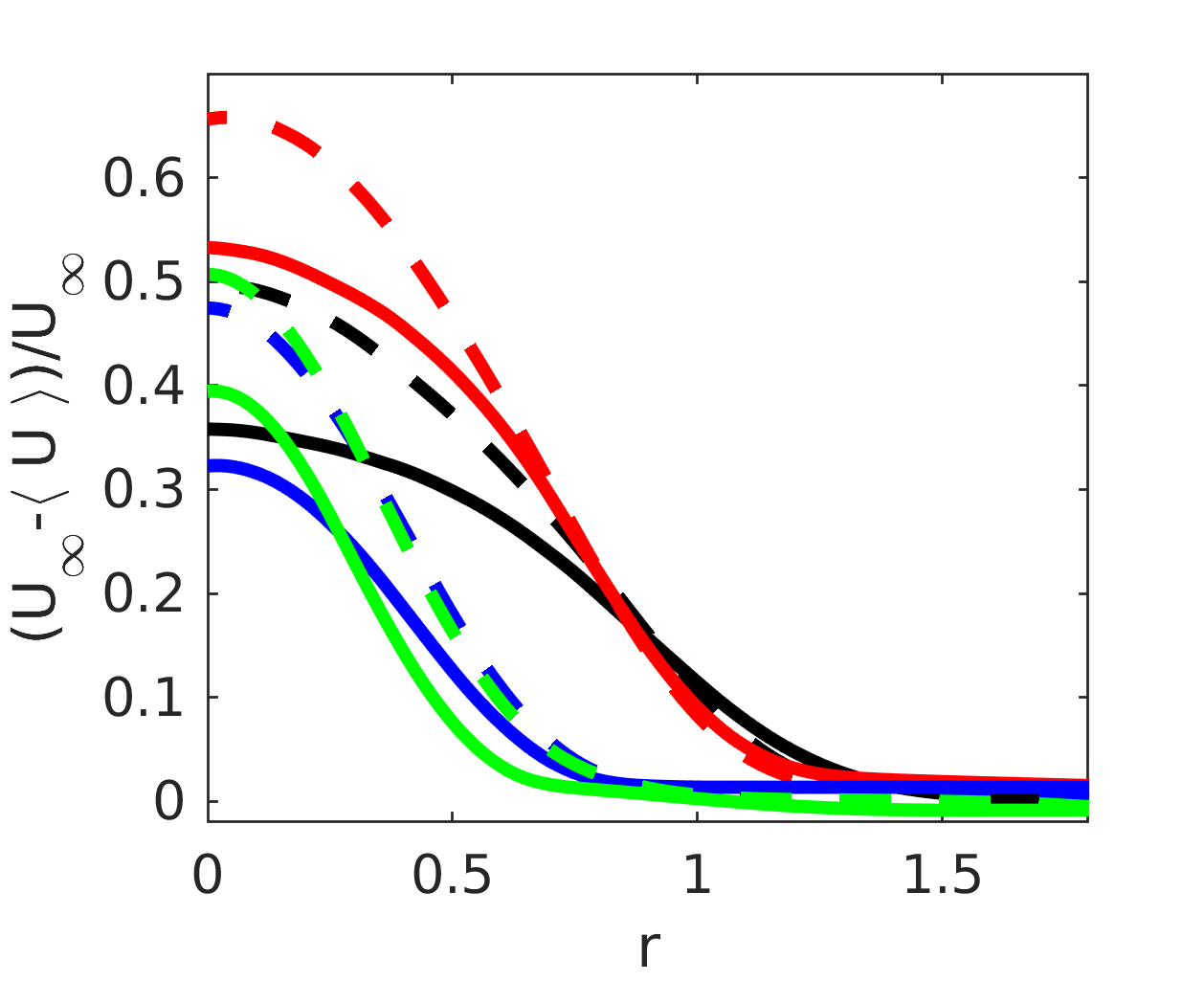}
   \includegraphics[width=5cm]{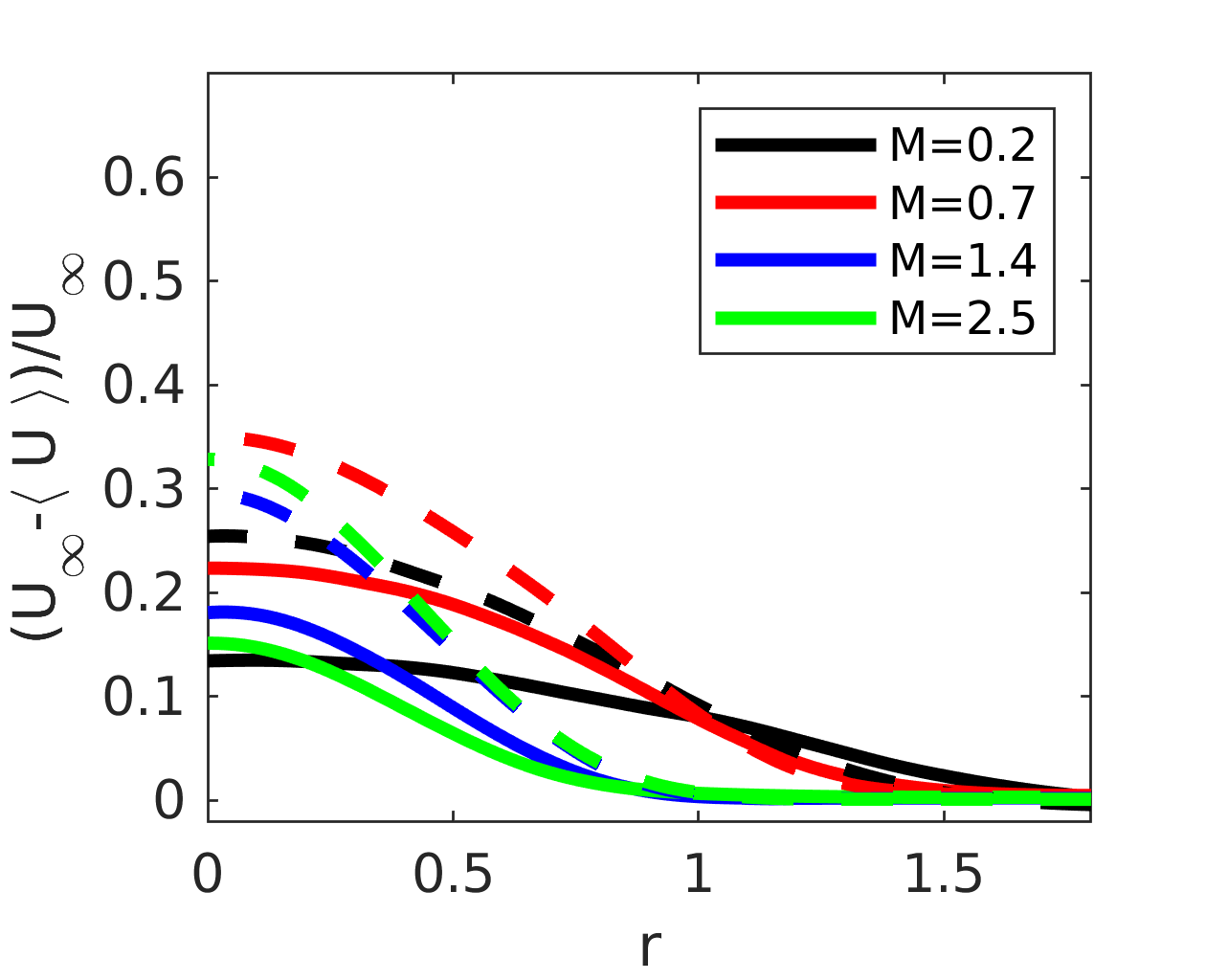}\\
   \hspace{0.35cm} (a) \hspace{4.5cm} (b)\\
 \end{center}
  \caption{\label{} Mean streamwise velocity component profile in the radial direction: a) x=5, b) x = 10; square plate (solid line) and fractal plate (dashed line).}
  \label{f10}
\end{figure}

In this section, we analyze the mean streamwise velocity component, Reynolds stress components, and TKE distributions in the crossflow, and along the centerline directions. The radial distribution of the mean flow velocity deficit $(U_{\infty} - \langle U \rangle)/U_{\infty}$ at streamwise locations $x=5$ and $x=10$ are plotted in figure \ref{f10} for all Mach numbers. For all subsequent figures, results corresponding to the square plate are represented by solid lines, while those corresponding to the fractal plate by dashed lines. For all cases, and at both streamwise locations, the higher velocity deficit for the fractal plate implies that the mean velocity in the wake is lower. The lowest scaled mean velocity (or the highest velocity deficit) is achieved at the high subsonic Mach number of $0.7$ (red curves in figure \ref{f10}), while the highest scaled velocity (or lowest velocity deficit) corresponds to the two supersonic cases, where both are in the same order of magnitude. The spread rate of the wake in the subsonic regime is larger than the supersonic regime, where it seems to decrease with increasing Mach number. The top parts of figures \ref{f4} and \ref{f5}, in fact, show that the wakes in the supersonic regime experience a size reduction in the radial direction up to approximately $x=2.5$ for $M=1.4$ and $x=4$ for $M=2.5$.

\begin{figure}
 \begin{center}
   \includegraphics[width=8cm]{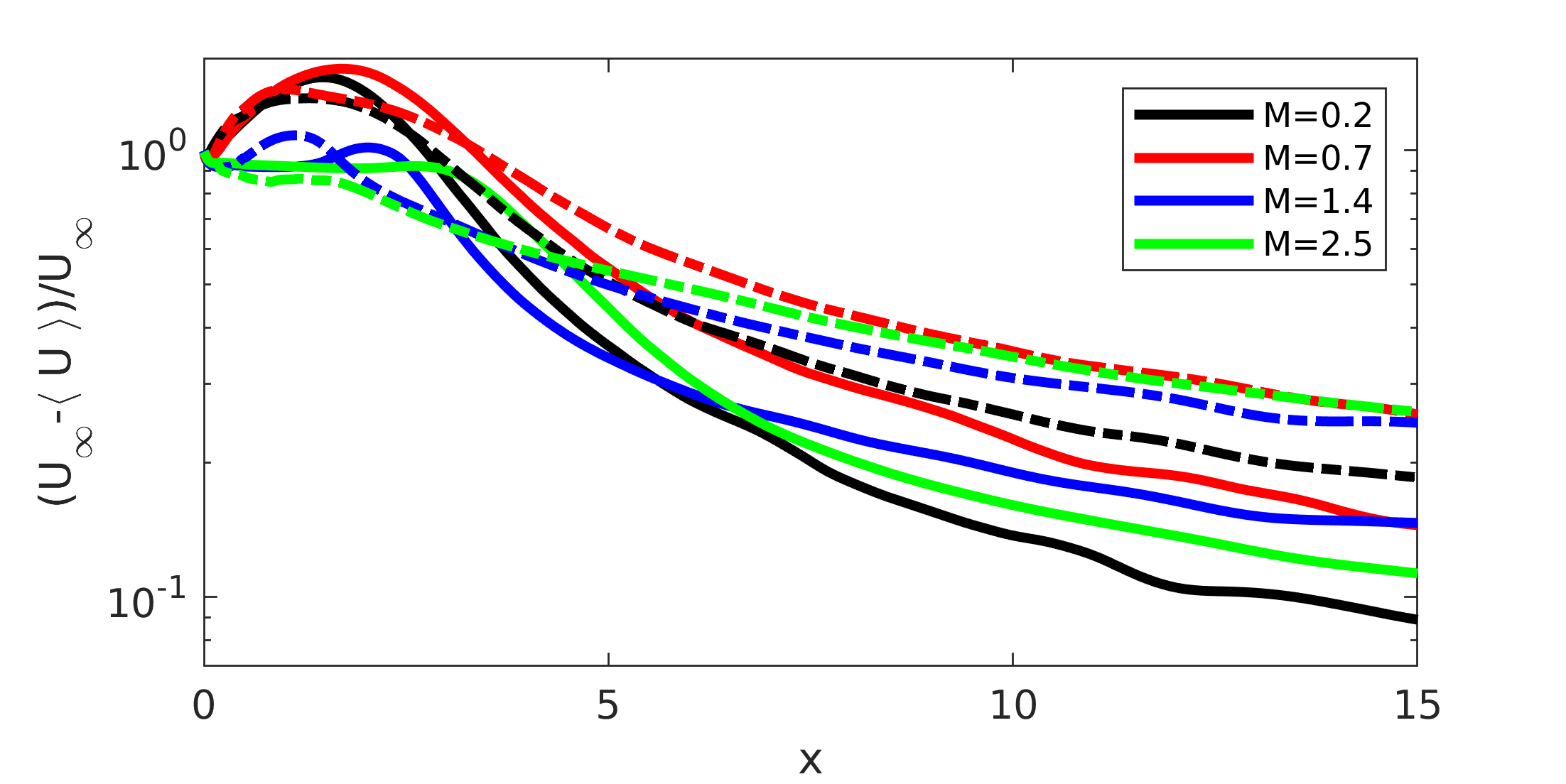}
 \end{center}
  \caption{\label{} Mean streamwise velocity component distribution along the streamwise direction; square plate (solid line) and fractal plate (dashed line).}
  \label{f11}
\end{figure}

Figure \ref{f11} illustrates the mean velocity deficit distribution along the centerline for all cases in logarithmic scale on the vertical axis. The same level of flow reversal, identified by regions of velocity deficit greater than one, is observed in both subsonic cases, while in the supersonic regime there is a low level of flow reversal for the $M=1.4$ case and none for the highest Mach number of $2.5$. For $x>4$, the rate of decay corresponding to the fractal plate is lower than that corresponding to the square plate, suggesting {\color{black} that the waves extend for longer distances in the downstream. This is a consequence of the increased mixing resulting from the interaction of the small fractal structures with the flow. At higher Mach numbers, the mixing in the wake is more intense due to the compressibility effects}.

\begin{figure}
 \begin{center}
   \hspace{-0.5cm}\includegraphics[width=4.13cm]{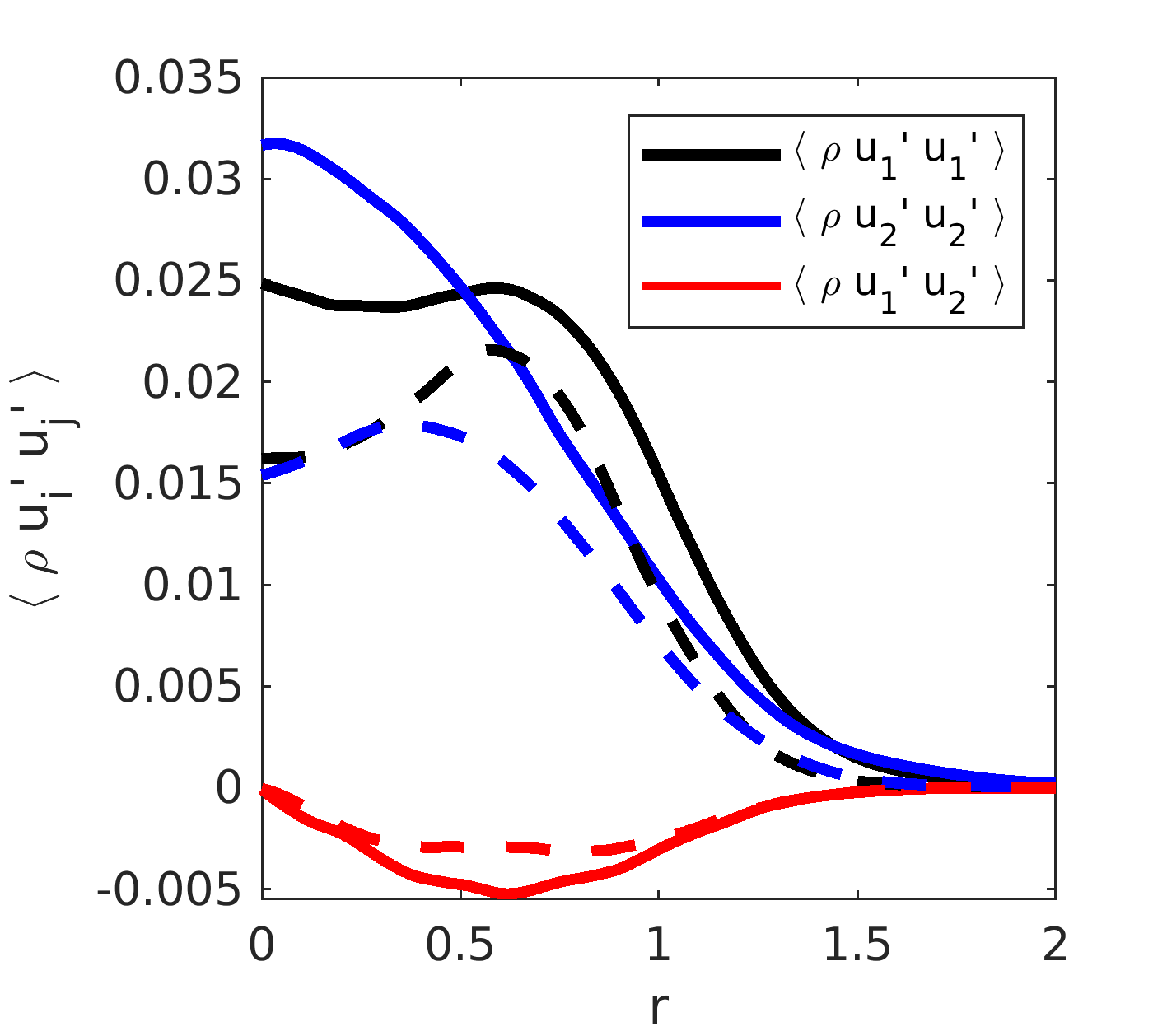}
   \includegraphics[width=4.13cm]{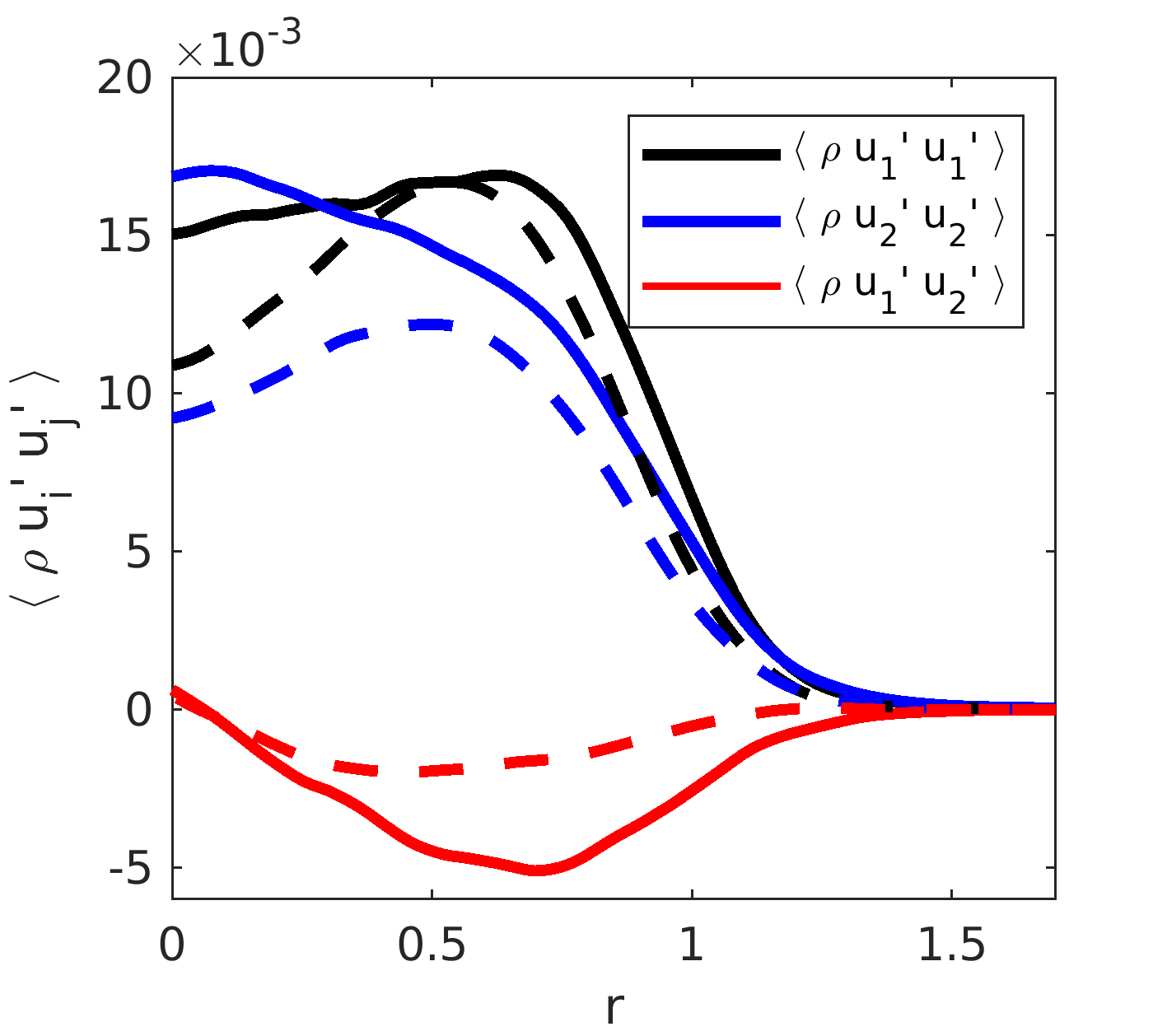}
   \includegraphics[width=4.13cm]{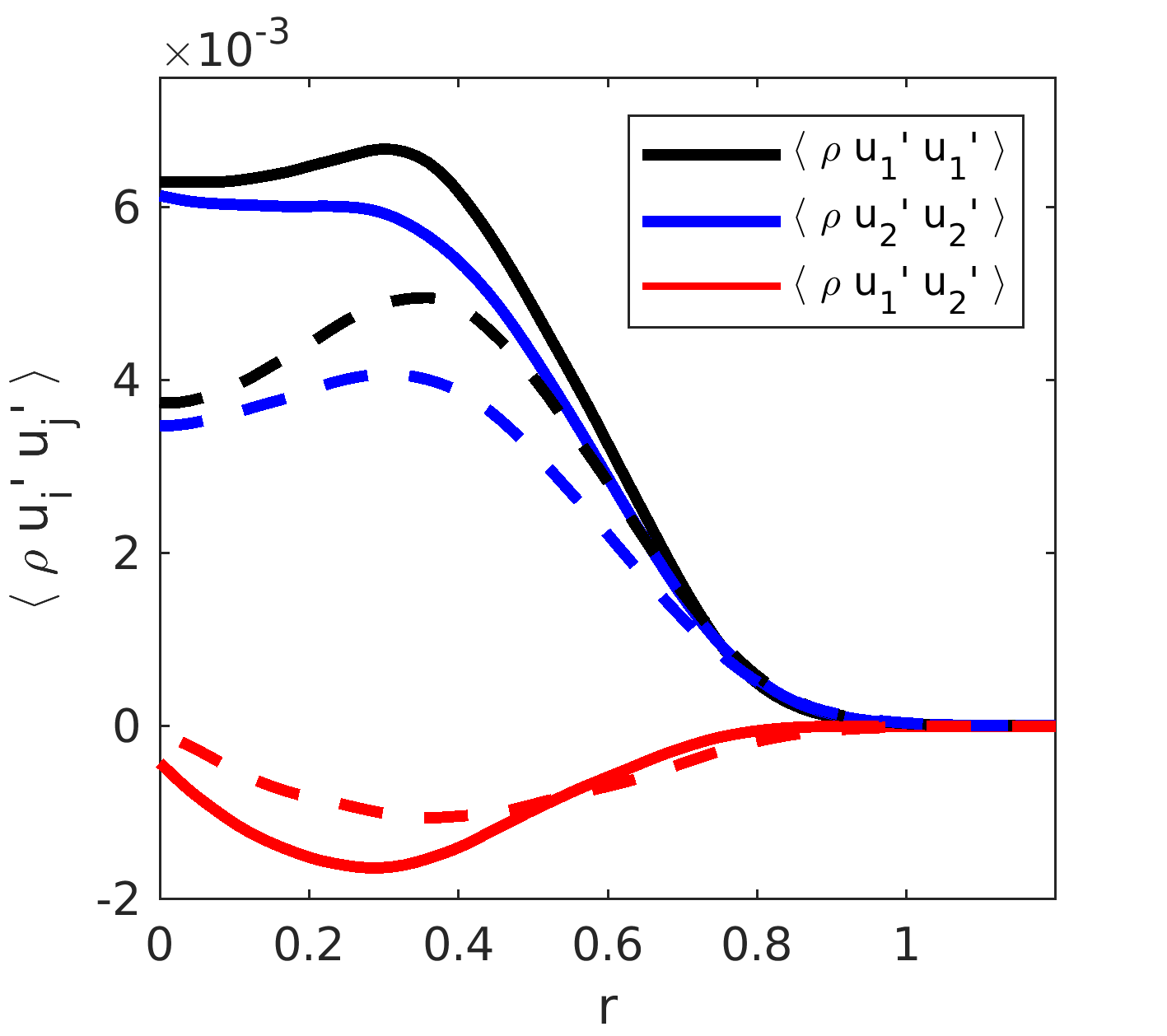}
   \includegraphics[width=4.13cm]{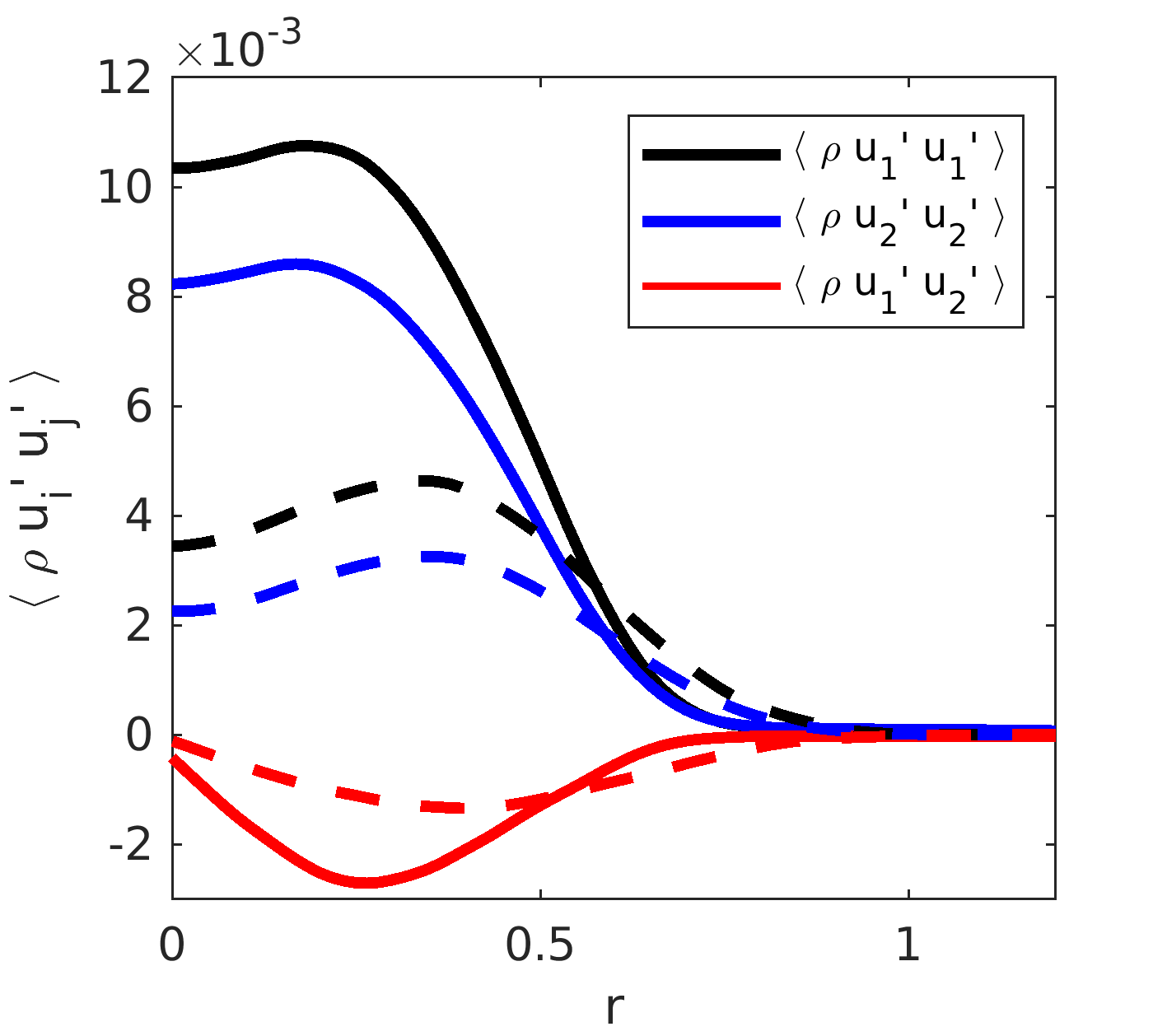}  \\
  \hspace{-0.5cm} \includegraphics[width=4.13cm]{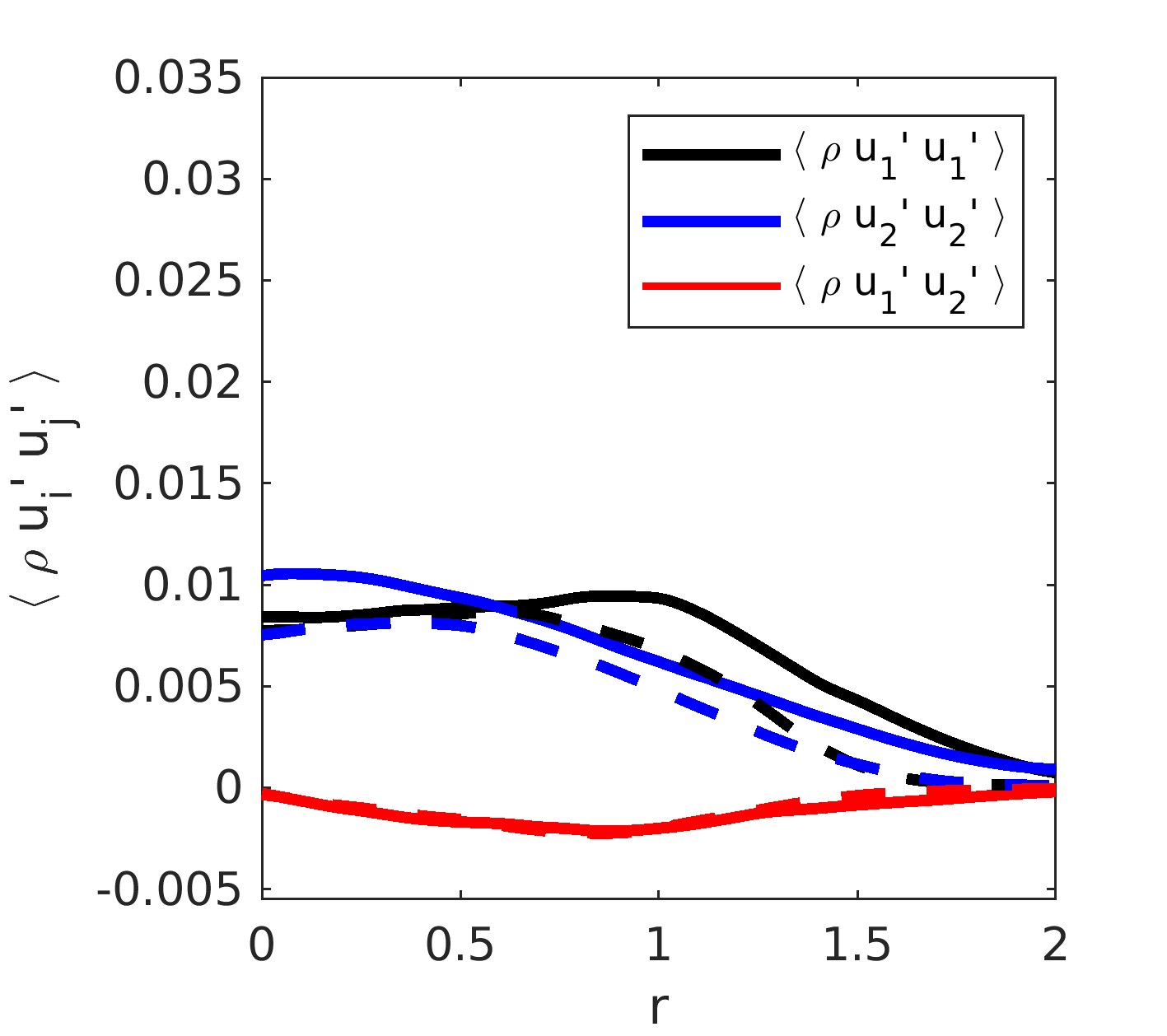}
   \includegraphics[width=4.13cm]{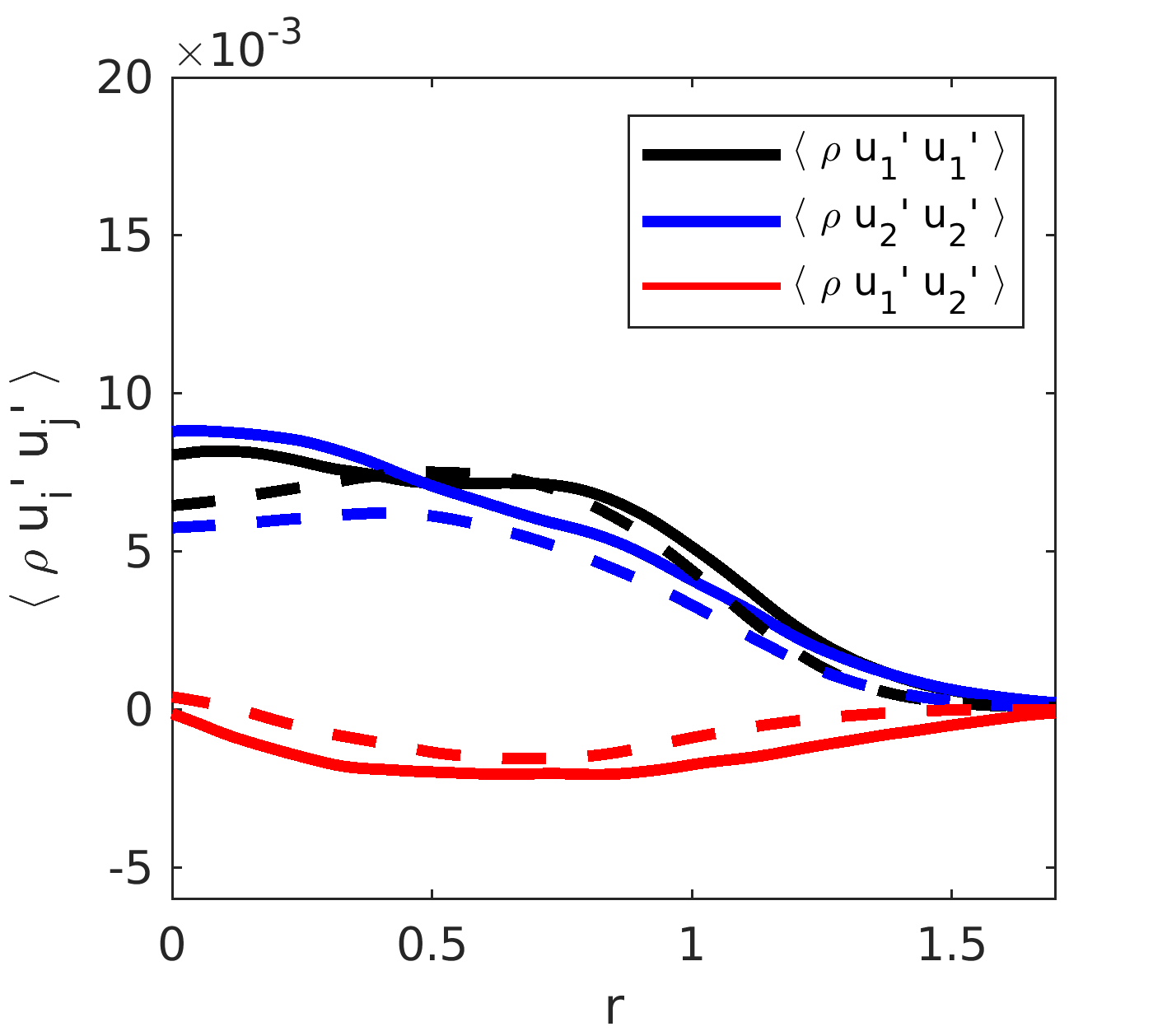} 
   \includegraphics[width=4.13cm]{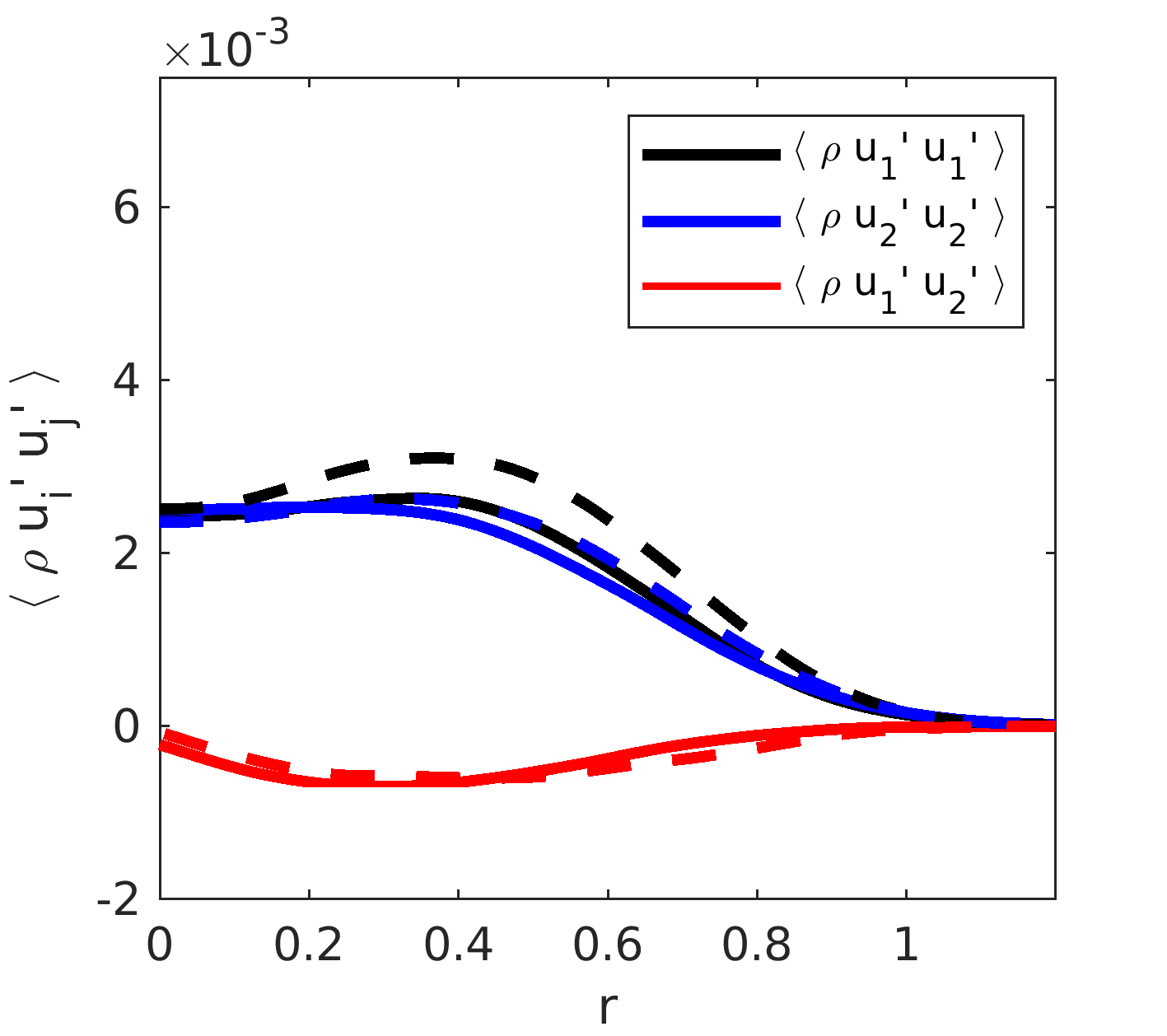}
   \includegraphics[width=4.13cm]{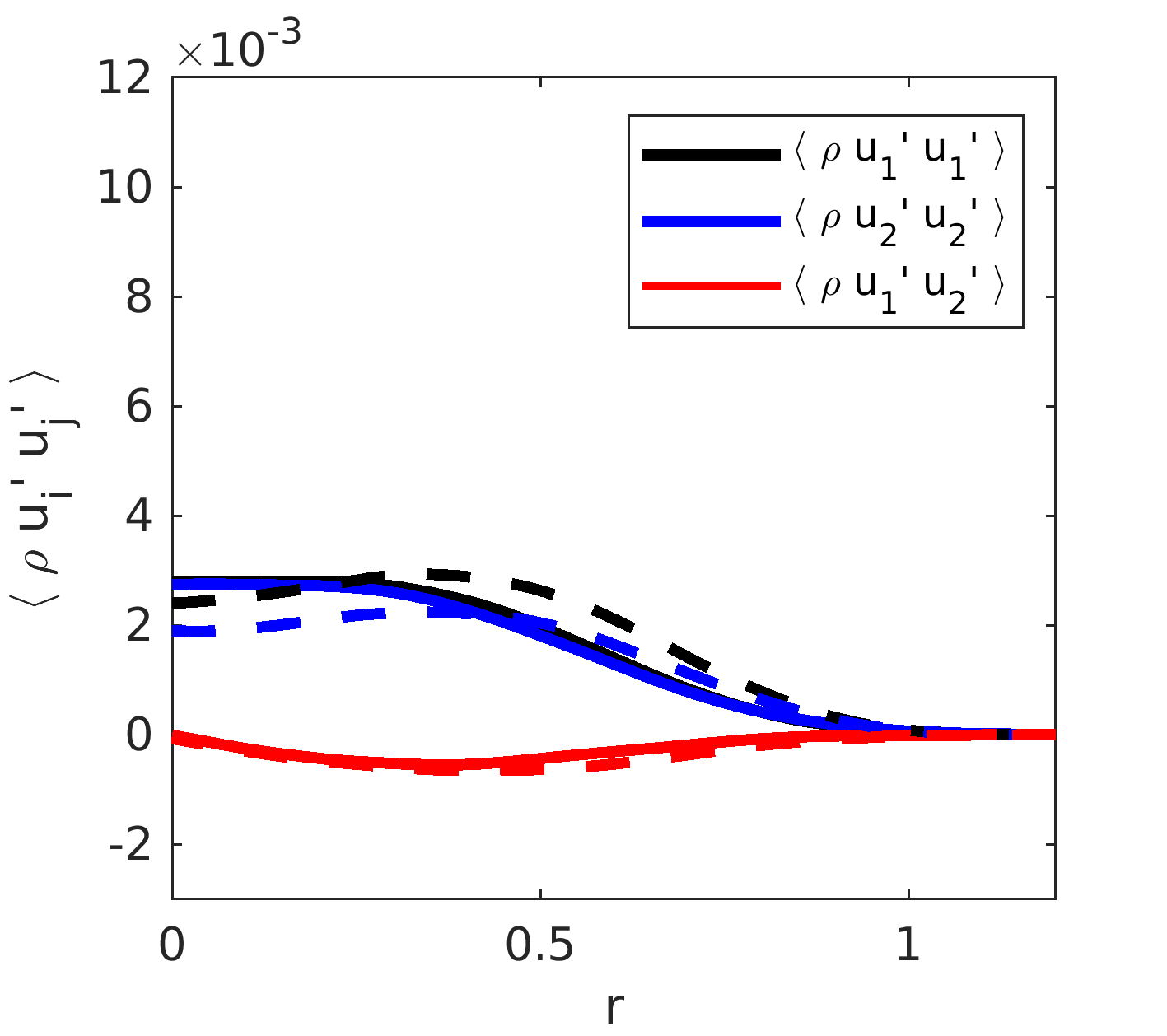}  \\
   (a) \hspace{36mm} (b) \hspace{36mm} (c) \hspace{36mm} (d)
 \end{center}
  \caption{\label{} Reynolds stress tensor components $\langle \rho u_1'u_1' \rangle$, $\langle \rho u_2'u_2' \rangle$, and $\langle \rho u_1'u_2' \rangle$ at $x=5$ (top row) and $x = 10$ (bottom row): a) $M=0.2$; b) $M=0.7$; c) $M=1.4$; d) $M=2.5$. Square plate are shown in solid line, and fractal plate in dashed line.}
  \label{f12}
\end{figure}

Reynolds stress tensor components $\langle \rho u_1'u_1' \rangle$, $\langle \rho u_2'u_2' \rangle$, and $\langle \rho u_1'u_2' \rangle$ were calculated by taking time averages. All three components of each case are superposed in the same graph as illustrated in figure \ref{f12}, at the streamwise locations $x=5$ and $x=10$. For all Mach numbers, these figures indicate that the $\langle \rho u_1'u_1' \rangle$ and $\langle \rho u_2'u_2' \rangle$ components are in the same order of magnitude for both square and fractal plates, while the shear stress component $\langle \rho u_1'u_2' \rangle$ is at a lower order of magnitude. Fractal plates appear to reduce the level of all Reynolds stresses when compared to results from the square plates at $x=5$; at $x=10$, however, they become comparable, and in the supersonic regime the level corresponding to the fractal plate exceeds that of the square plate.

\begin{figure}
 \begin{center}
   \includegraphics[width=8cm]{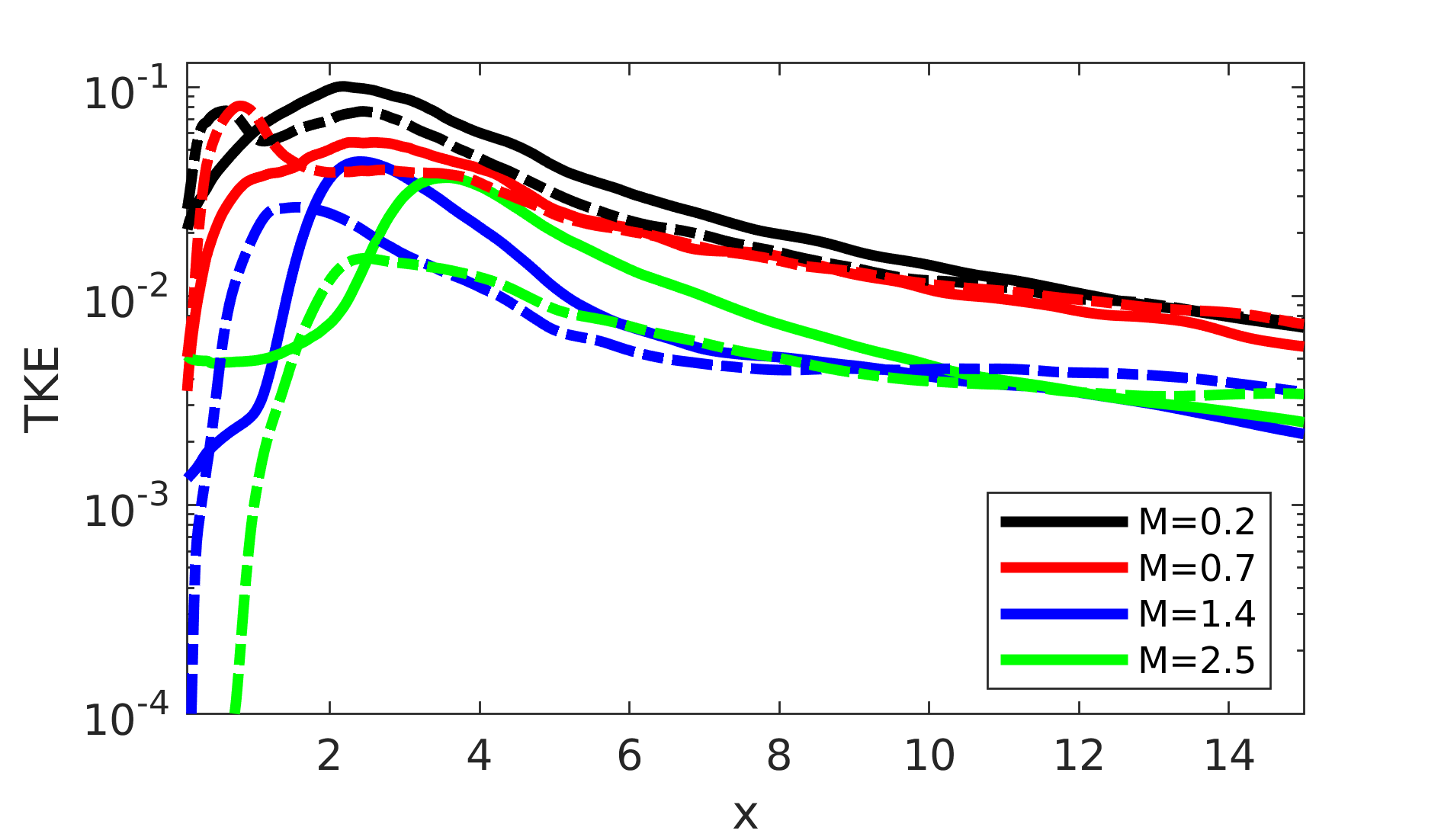}
 \end{center}
  \caption{\label{} TKE distribution along the streamwise direction in the shear layer; square plate (solid line) and fractal plate (dashed line).}
  \label{f13}
\end{figure}

Figure \ref{f13} displays scaled TKE along the shear layer for all cases. The two curves corresponding to the smallest Mach number ($0.2$) attain the maximum roughly at the same streamwise location. As the Mach number increases, the {\color{black}less intense} maximum TKE corresponding to the fractal plate moves upstream. {\color{black}In the supersonic regime, we notice that the turbulence intensity of the square plates is small downstream of the plate for a longer distance (approximately $1$ square side for $M=1.4$ and $2$ square sides for $M=2.5$).}

\subsection{Two-point velocity correlations}\label{}
  
\begin{figure}
 \begin{center}
   \hspace{-0.5cm}\includegraphics[width=4.13cm]{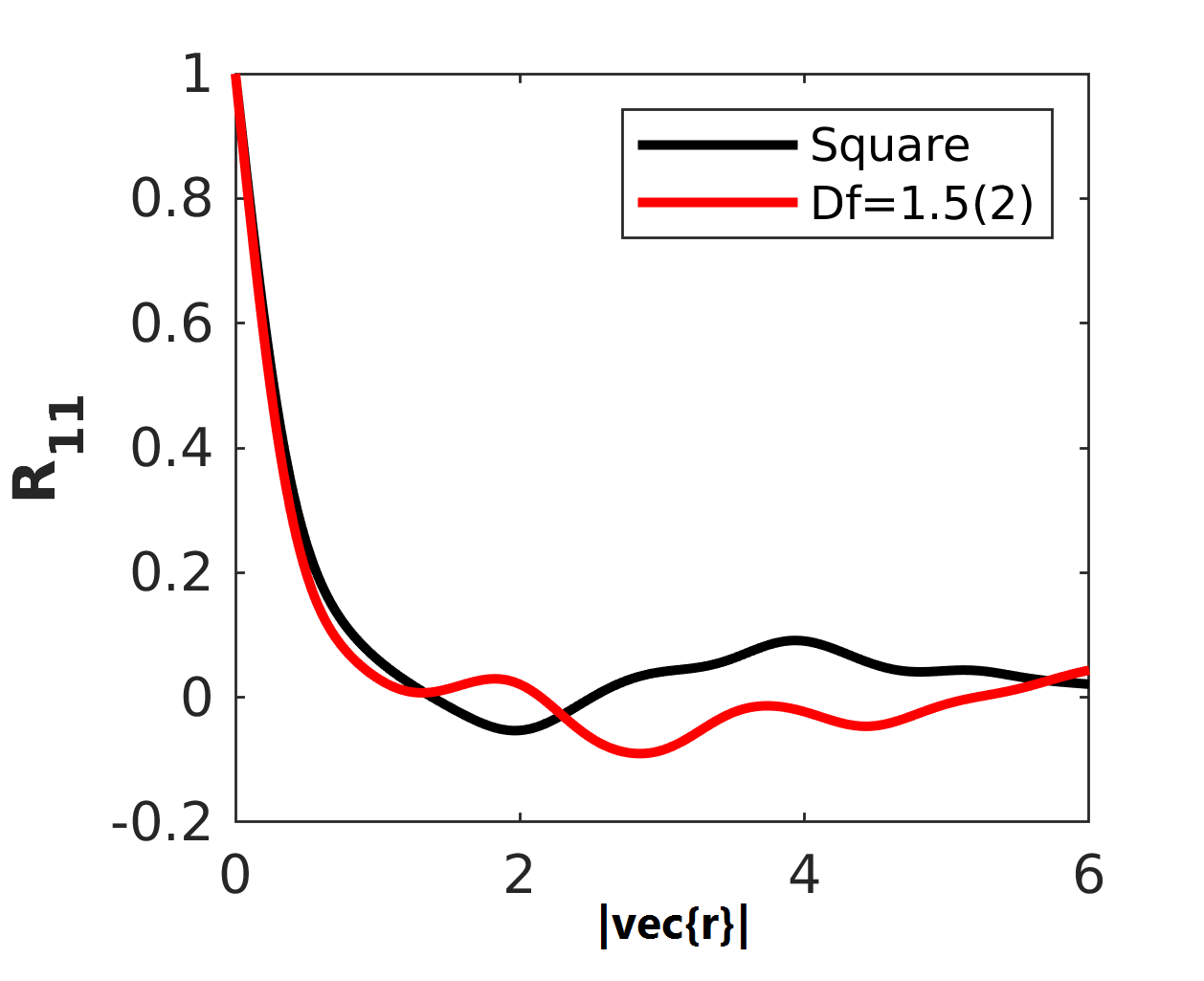}
   \includegraphics[width=4.13cm]{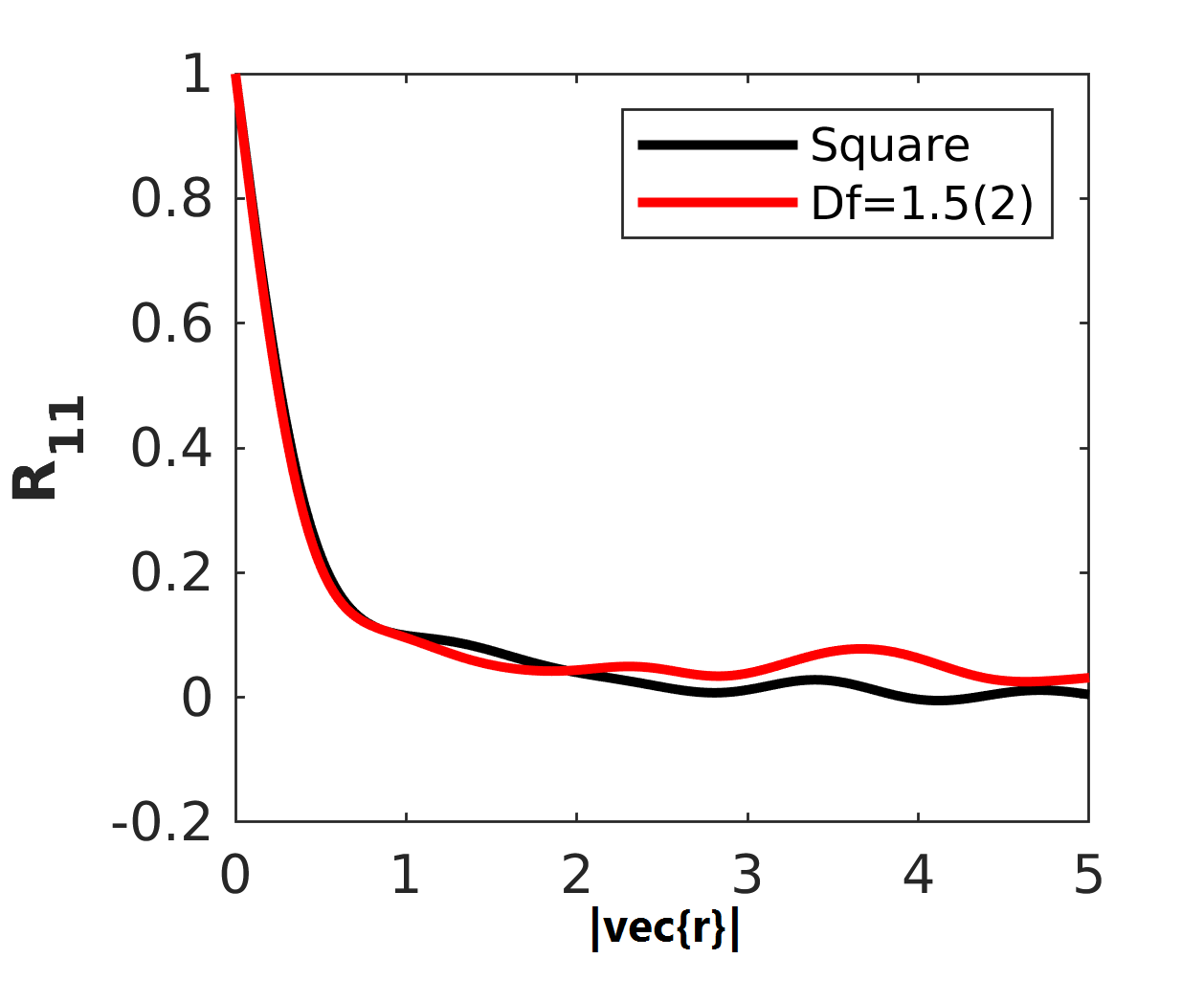}
   \includegraphics[width=4.13cm]{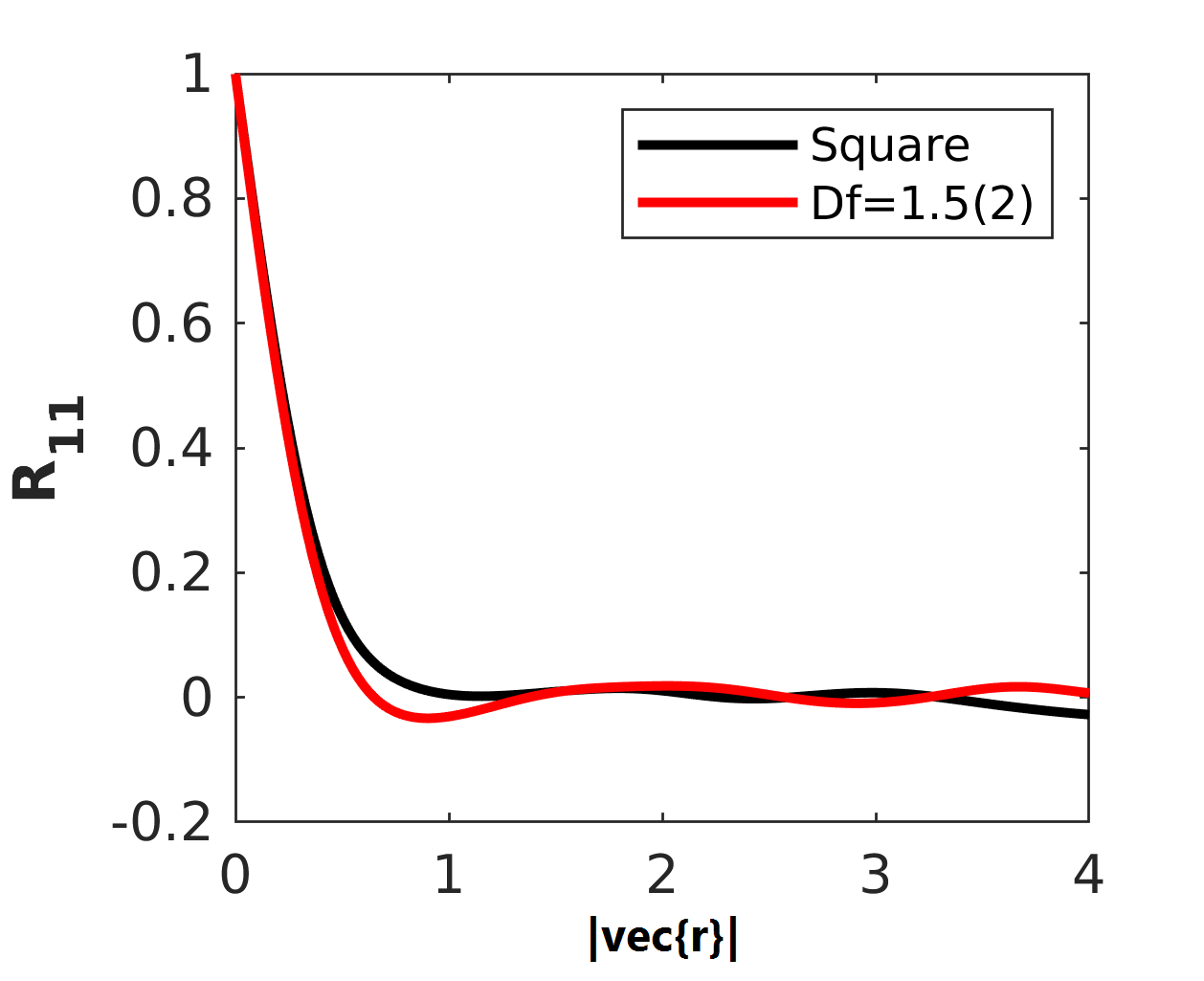}
   \includegraphics[width=4.13cm]{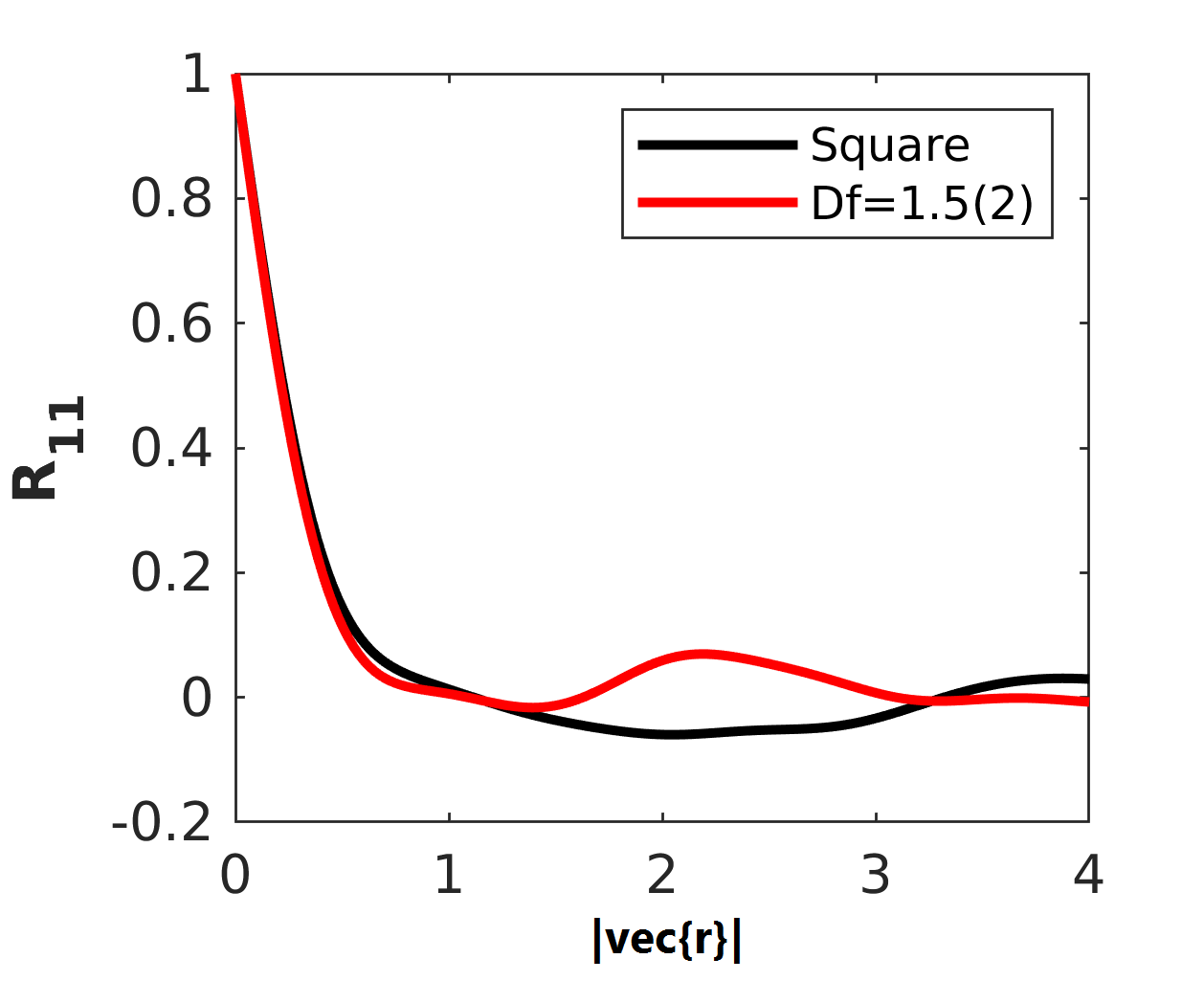}  \\
  \hspace{-0.5cm} \includegraphics[width=4.13cm]{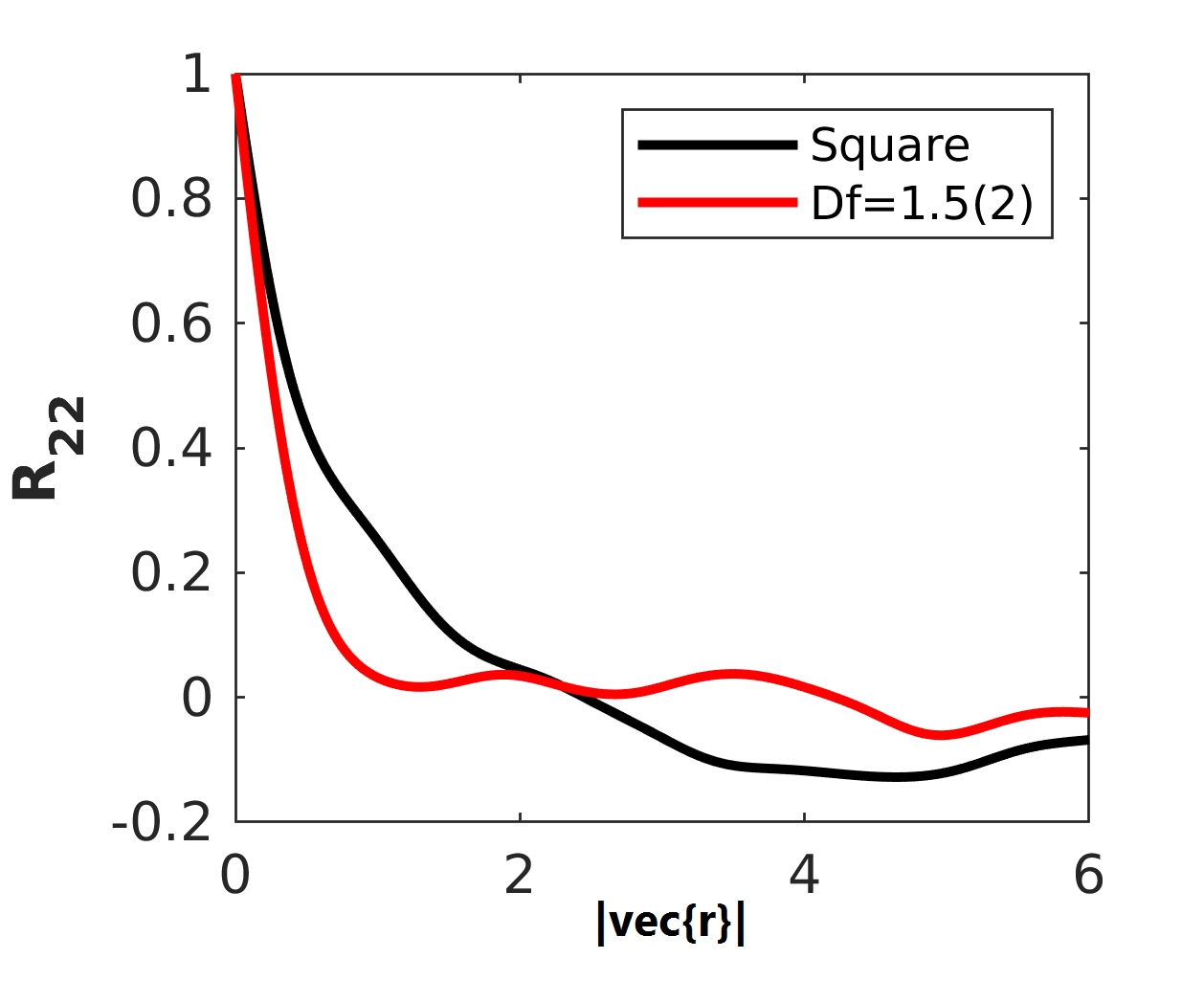}
   \includegraphics[width=4.13cm]{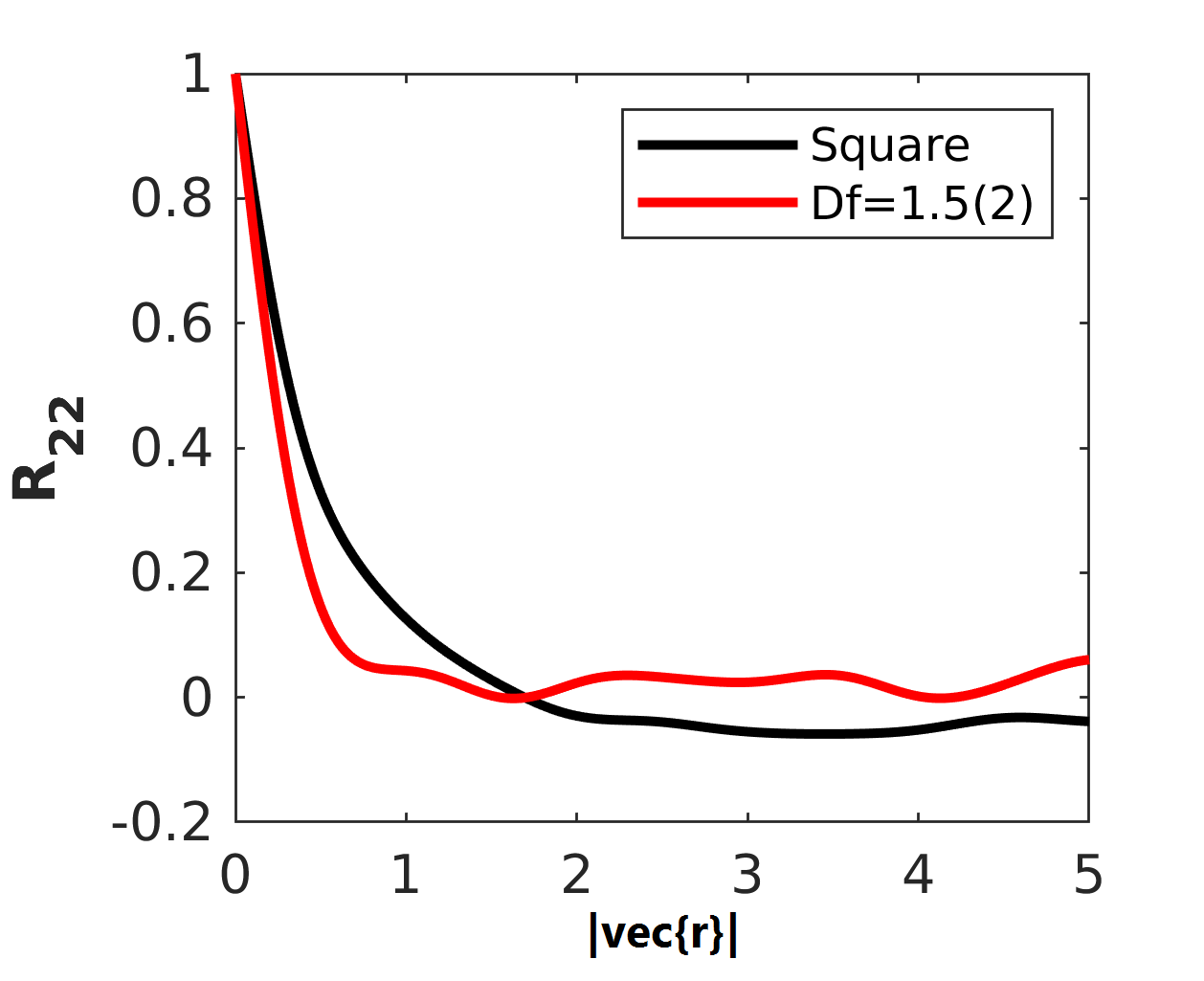} 
   \includegraphics[width=4.13cm]{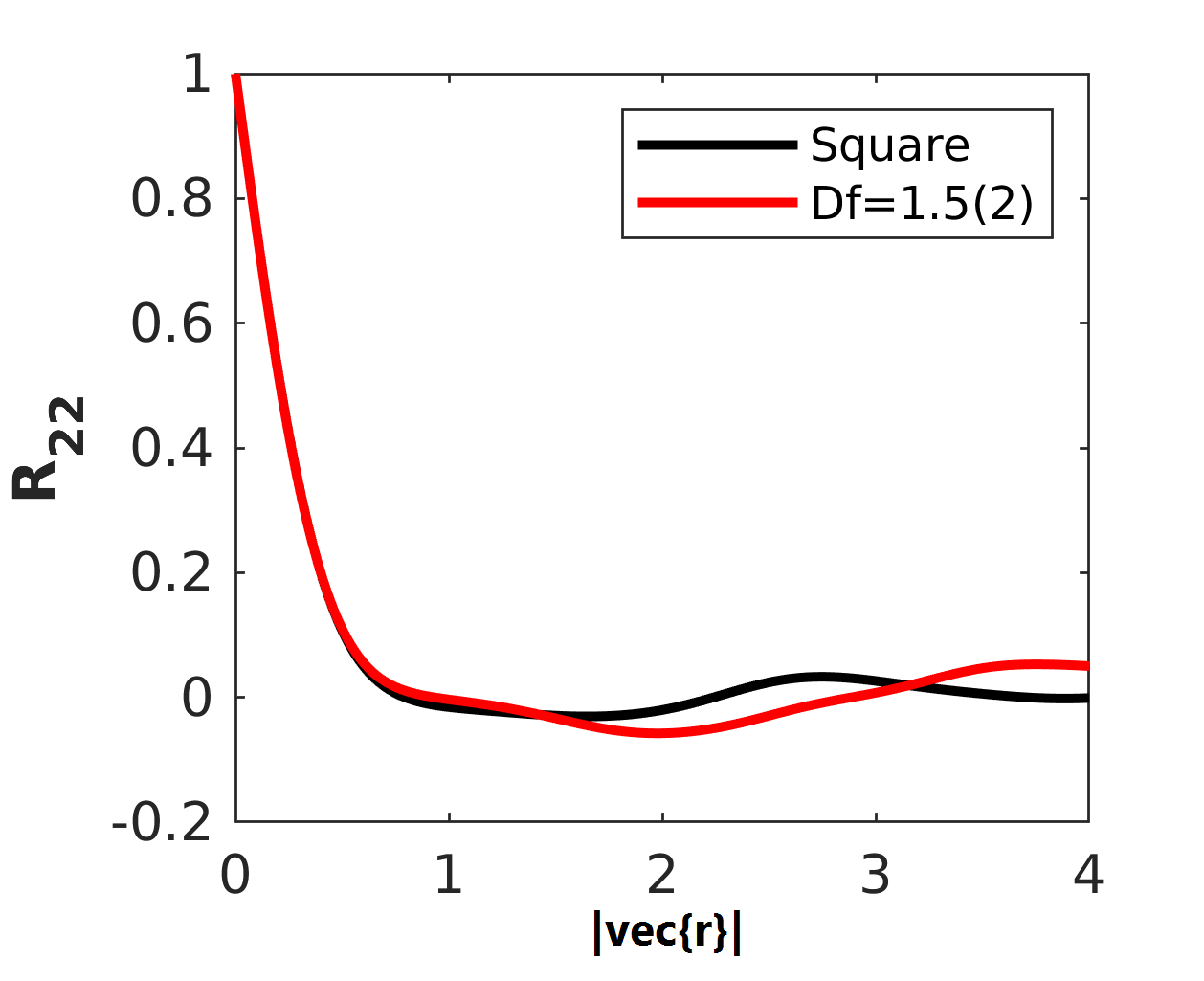}
   \includegraphics[width=4.13cm]{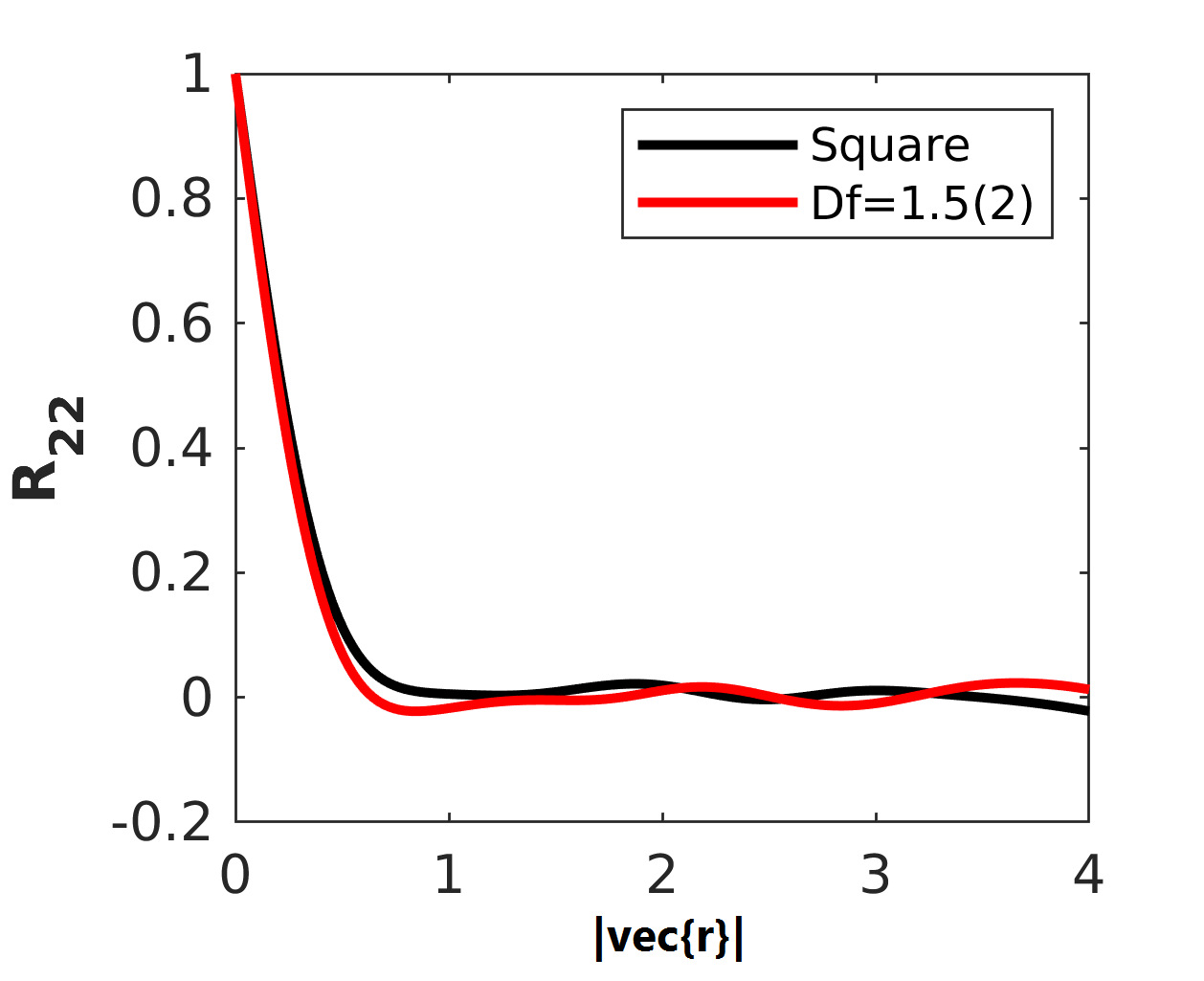}  \\
   (a) \hspace{36mm} (b) \hspace{36mm} (c) \hspace{36mm} (d)
 \end{center}
  \caption{\label{} Longitudinal $R_{11}$ (top row), and transverse $R_{22}$ (bottom row) two-point velocity correlation at $x=3$ ( a $\&$ b) and $x=5$ (c $\&$ d): a) $M=0.2$; b) $M=0.7$; c) $M=1.4$; d) $M=2.5$.}
  \label{f14}
\end{figure}

\begin{figure}
 \begin{center}
    \hspace{-0.5cm}\includegraphics[width=4.13cm]{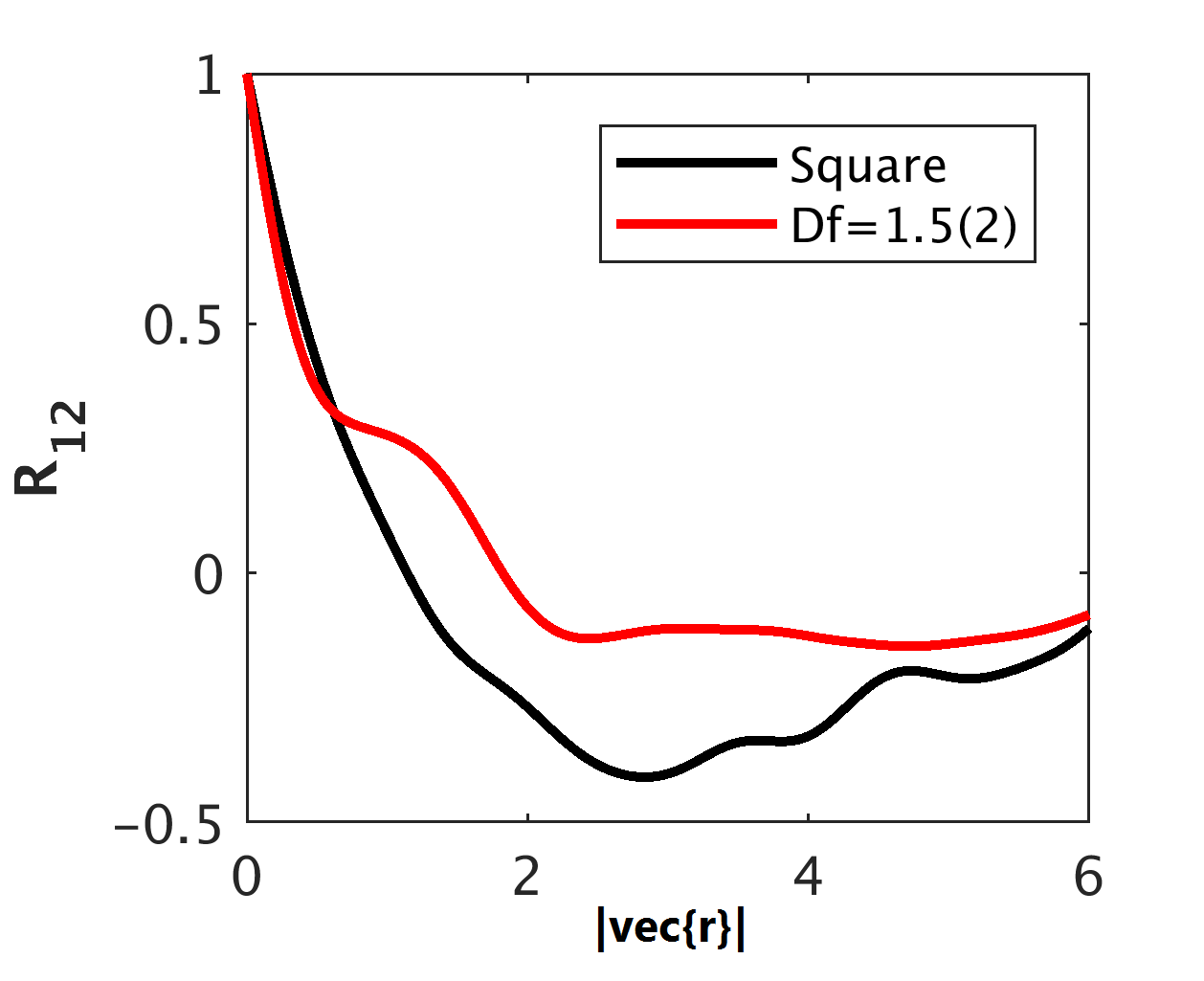}
   \includegraphics[width=4.13cm]{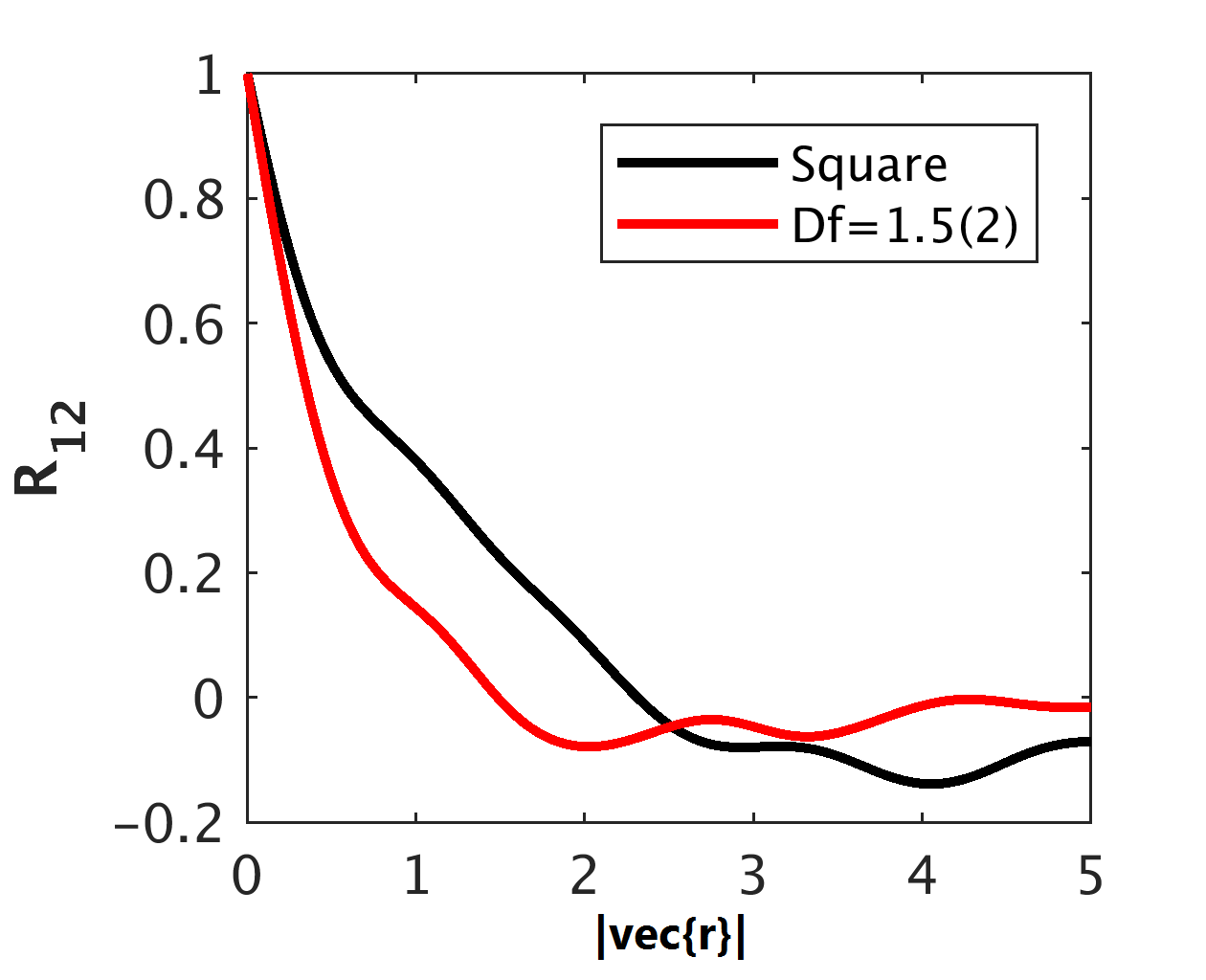} 
   \includegraphics[width=4.13cm]{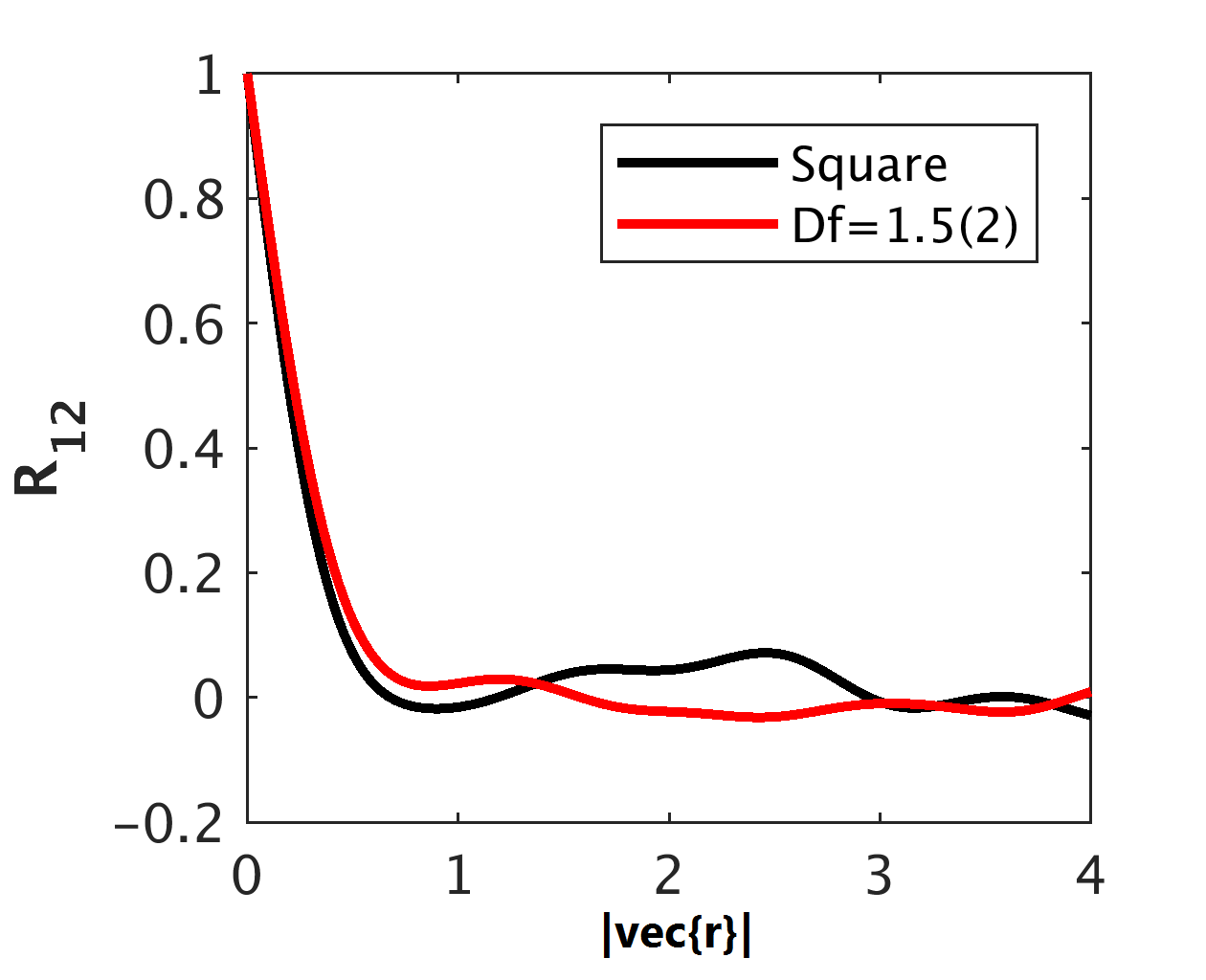}
   \includegraphics[width=4.13cm]{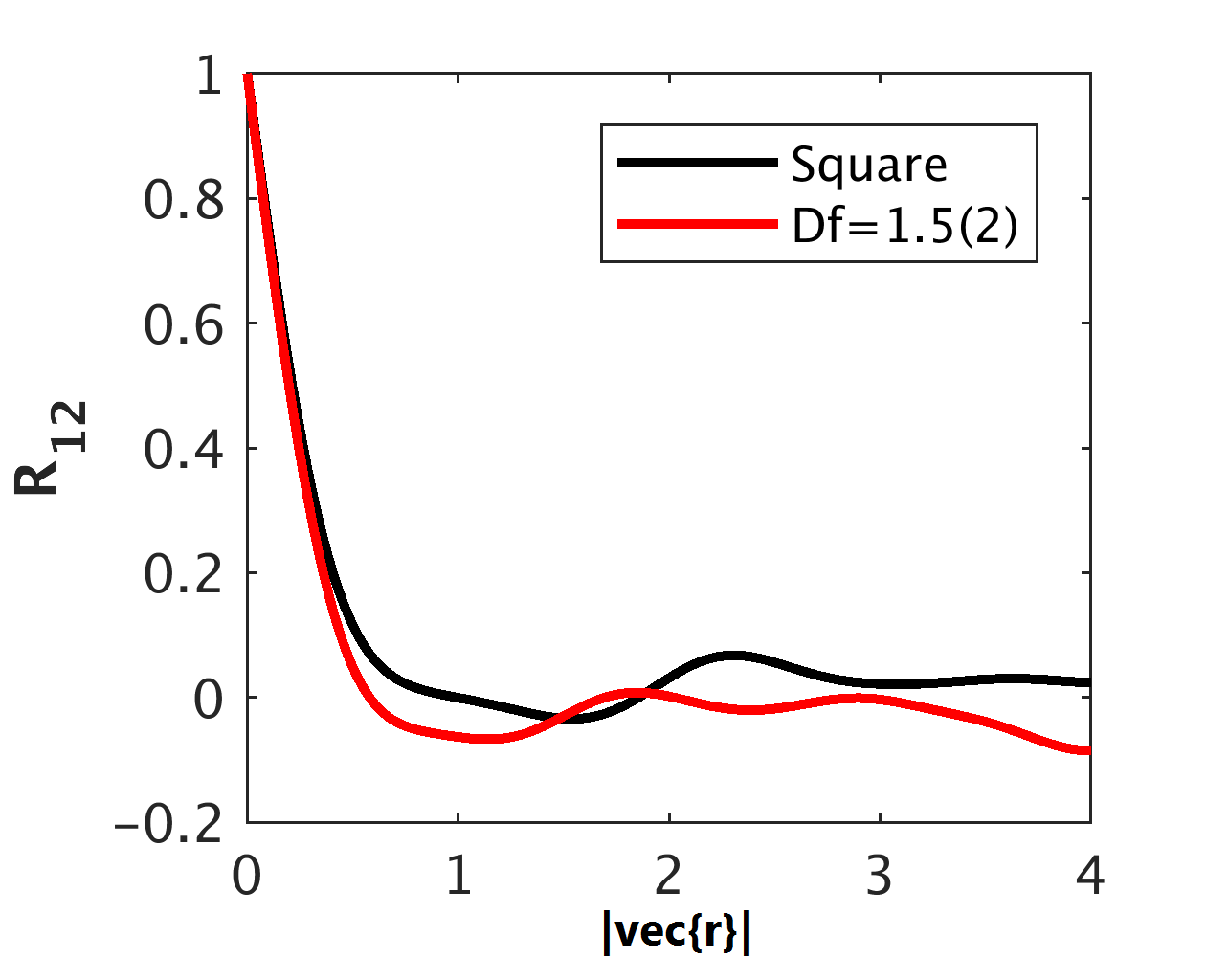}  \\
   (a) \hspace{3.5cm} (b)  \hspace{3.5cm} (c) \hspace{3.5cm} (d)
 \end{center}
  \caption{\label{} Deviatoric $R_{12}$ two-point velocity correlation at $x=3$ ( a $\&$ b) and $x=5$ (c $\&$ d): a) $M=0.2$; b) $M=0.7$; c) $M=1.4$; d) $M=2.5$.}
  \label{f15}
\end{figure}

Following Batchelor's notation (p. 24 of Batchelor\cite{Batchelor}), we define the velocity correlation tensor for two points separated by the space vector $\mathbf{r}$ as follows
\begin{equation}
 R_{ij}(\mathbf{r},\mathbf{x},t) = \langle u_i'(\mathbf{x},t) u_j'(\mathbf{x}+\mathbf{r},t) \rangle
\end{equation}
where $u_i'$ are velocity fluctuations, and $\bf{r}=r\hat{i}$ is the vector separation between the two correlated points in the x-direction. We analyze the longitudinal ($R_{11}$), transverse ($R_{22}$), and deviatoric ($R_{12}$) correlation tensor components along the centerline ($y=z=0$), starting at a certain streamwise location ($x=3$ for the subsonic cases, and $x=5$ for the supersonic cases), by taking a time average over a long interval of time (2 to 3 shedding cycles). Figures \ref{f14} and \ref{f15} show these correlations for all considered Mach numbers. Wake meandering is clearly evident in the $R_{22}$ correlation at $M=0.2$ (see figure \ref{f14}a, right), via negative (or, anti-) correlation at large separation, $r$ (in other words, the transverse velocity component undergoes sign change because of this meandering effect). 

We observe further influence of meandering in the transverse correlation $R_{22}$ at $M=0.7$ (as figure \ref{f14}b indicates, right). It is interesting to note that there is not much difference between the correlations corresponding to the square and fractal plates for the supersonic cases (figures \ref{f14}c and d).
This indicates that the size of the unsteady flow structures (quantified by the integral length scale, estimated by integrating the appropriate correlation function) generated by both plates does not vary significantly.

While the fluid dynamics of wakes and jets differ in terms of the streamwise spatial growth scale and temporal intermittency (p. 151 of Pope \cite{Pope}), the basic structure of the turbulence correlations in Figs \ref{f14} and \ref{f15} are similar to jets (see Fig. 17 in Pokora $\&$ McGuirk \cite{Pokora} and p. 202 of Townsend \cite{Townsend}) especially in the region far downstream from the fractal or square shape, away from the region where the wake flow periodically modulates. This similitude warrants a closer examination of the correlations in an attempt to explain potential aerodynamic noise reduction, which is more relevant and practical to jets.
The faster spatial decay of the deviatoric correlation tensor component, $R_{12}$, at lower flow speeds of $M = 0.2$ and $M = 0.7$ in figures \ref{f15}a and b for the fractal geoemetry clearly then has implications for aero-acoustic modeling. 
Figure 8 and Eqs. (6.21), (6.24), (6.27) $\&$ (7.13) in Goldstein $\&$ Leib \cite{Goldstein_and_Leib} shows that the noise radiated by a second-order correlation of the type $R_{12}$, dominates the peak jet noise.
The above correlations include density fluctuations  however (i.e. in the form of the density-weighted two-point time-delayed correlation) but the density effect is known to have a small or negligible impact on the aero-acoustics of isothermal jets (see Afsar \textit{et al}. \cite{Afsar1,Afsar2}). 
%
In Breda and Buxton \cite{Breda_and_Buxton_1,Breda_and_Buxton_2}, fractal jets were found to have more rapid de-correlation compared to those corresponding to the square jets at incompressible Mach numbers. Moreover, our findings indicate that this rapid de-correlation persists even for compressible Mach numbers in wakes. Therefore, these results serve as a strong evidence that fractals could ultimately reduce the peak jet noise in the compressible regime as well. 
This conclusion is consistent with the reduction in the pressure spectrum shown in figures \ref{f17}a and b.
%
%
Note that the effect of wake meandering could be one of the reasons for the relatively smaller `size' of the anti-correlation region in figures \ref{f14} $\&$ \ref{f15} for $R_{22}$ and $R_{12}$ respectively as compared to a jet flow for example (see Pokora $\&$ McGuirk \cite{Pokora}). It is worth noting however that the latter authors presented the temporal variation of the above second-order correlation functions, which should not be much different from the spatial correlation under the assumption that Taylor's frozen turbulence hypothesis approximately holds.

\subsection{Energy and pressure spectra}\label{}

\begin{figure}
 \begin{center}
  \hspace{-0.5cm} \includegraphics[width=4.13cm]{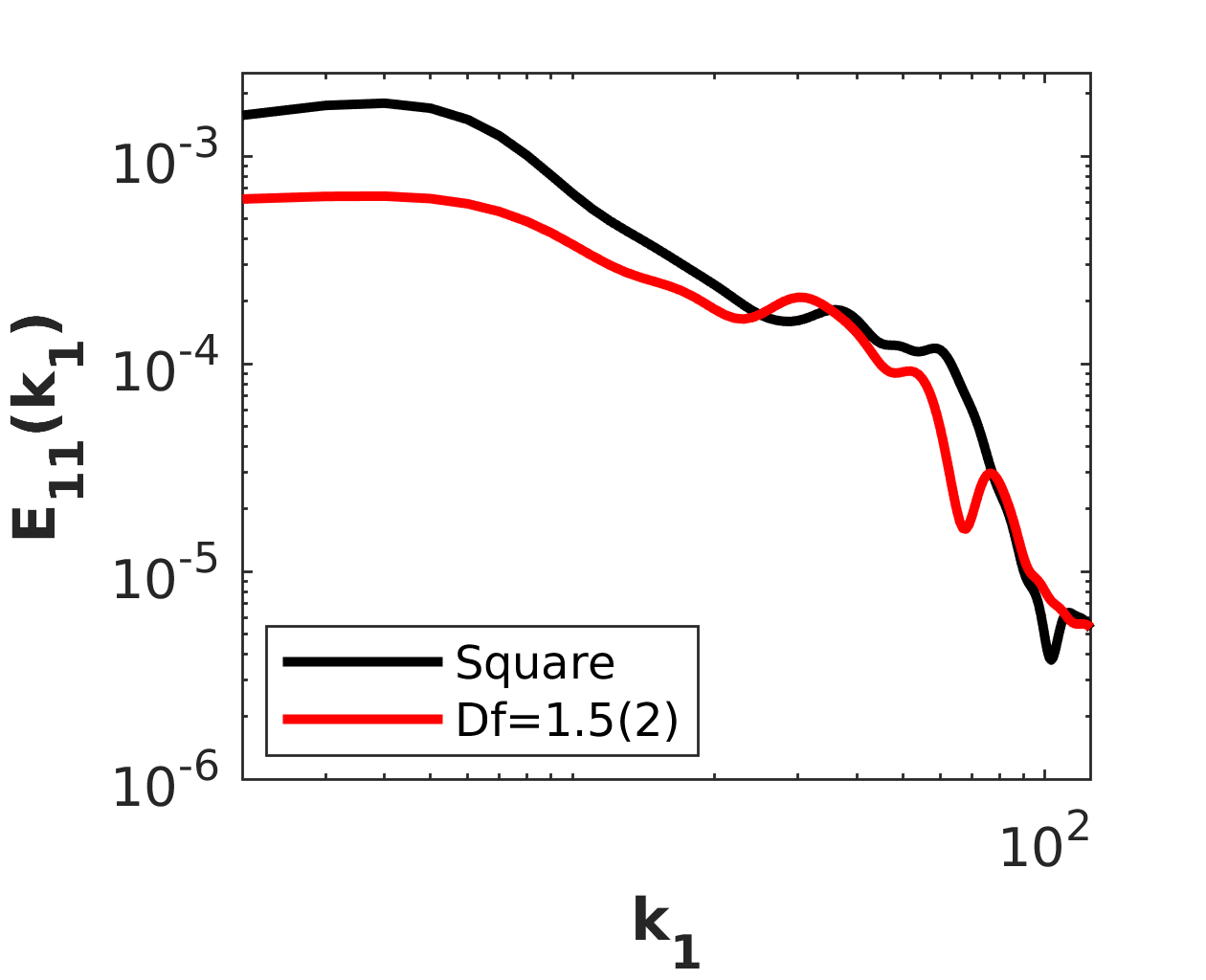}
   \includegraphics[width=4.13cm]{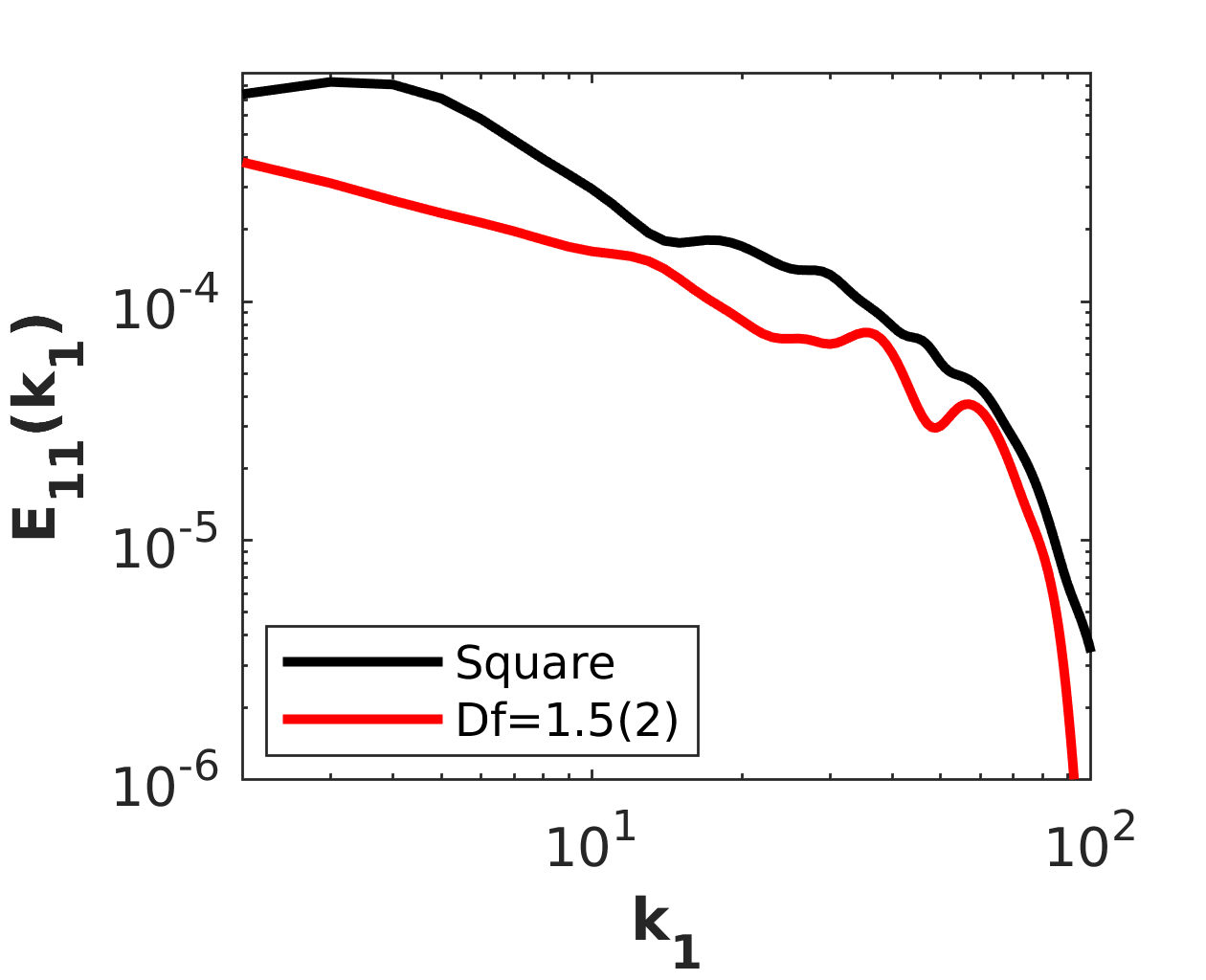}  
   \includegraphics[width=4.13cm]{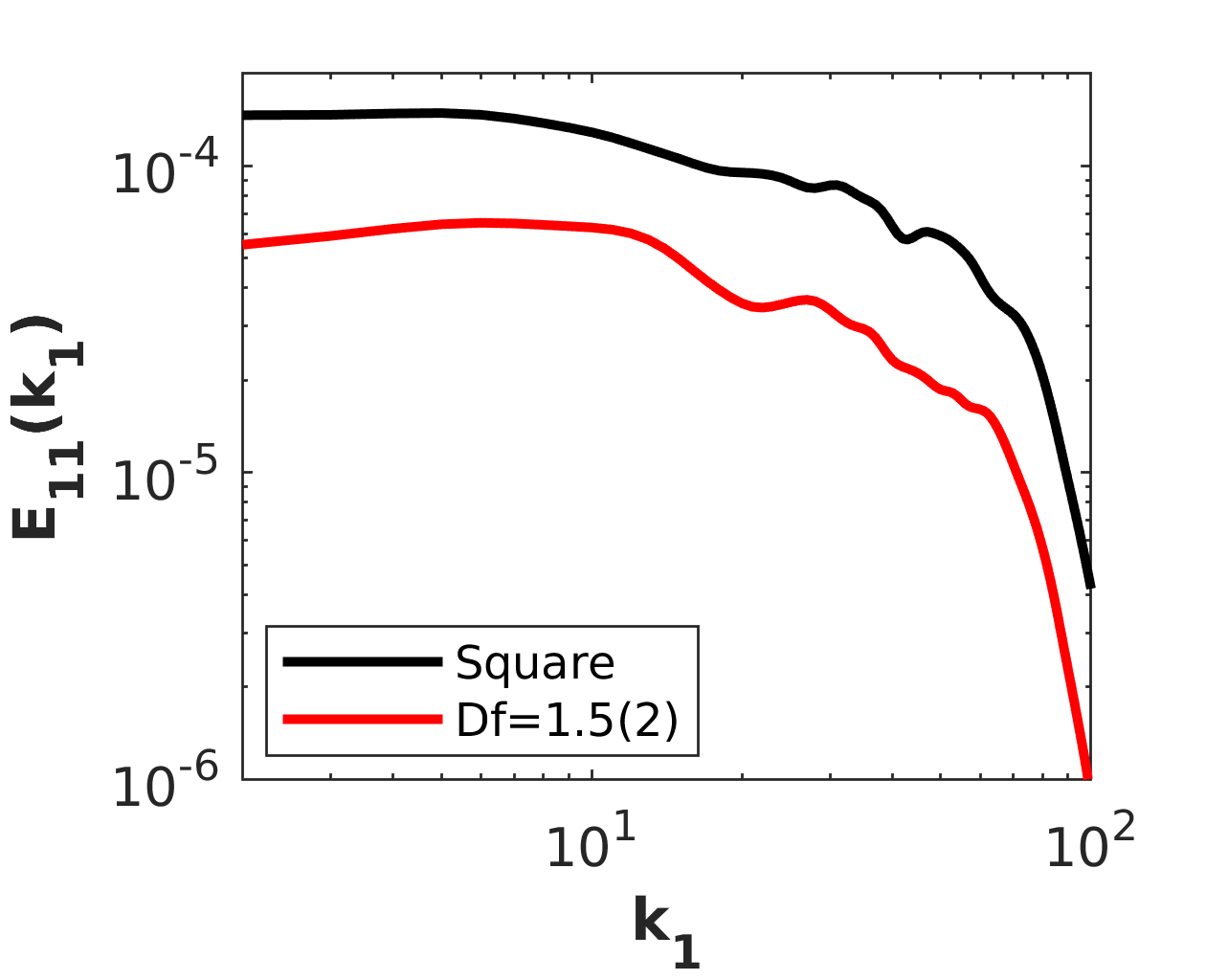}
   \includegraphics[width=4.13cm]{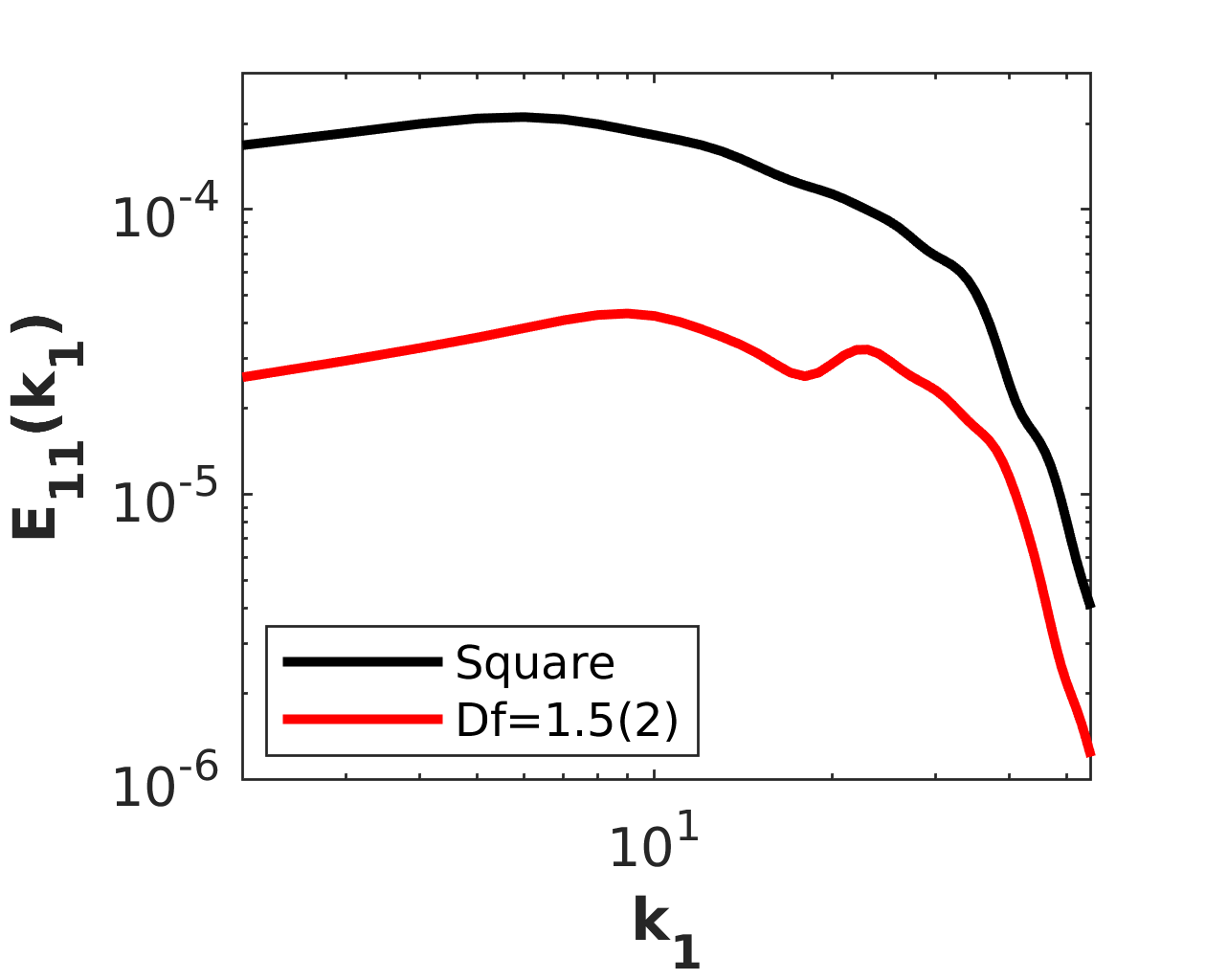}   \\
   (a) \hspace{3.5cm} (b)  \hspace{3.5cm} (c) \hspace{3.5cm} (d)
 \end{center}
  \caption{\label{} One-dimensional turbulence energy spectrum: a) $M = 0.2$; b) $M = 0.7$; c) $M = 1.4$; d) $M = 2.5$.}
  \label{f16}
\end{figure}

  The one-dimensional energy spectrum can be calculated as
\begin{equation}
 E_{ij}(k_1,t) = \frac{1}{\pi} \int_{-\infty}^{\infty} R_{ij}(\mathbf{e}_1 r_1)e^{-i k_1 r_1} dr_1
\end{equation}
where $k_1$ is the wavenumber component, and $\mathbf{e}_1$ is the unit vector along $r_1$. We plot the energy spectrum in the shear layer ($y=0.5$) and further downstream ($x=4$) in figure \ref{f16} for all Mach numbers, with black curves corresponding to the square plate, and red curves to the fractal plate. We can see that, for all Mach numbers, the energy of the fractal-generated wake is smaller than that contained in the square-generated wake, to a lesser extent for the large wavenumber range of the subsonic cases ($M=0.2$ and $M=0.7$). This result indicates that the fractal plate breaks up the large scale flow structures into much smaller, less energy-containing remnants, especially at supersonic speeds. This is because the unsteady flow scales at all wavenumbers are affected by the fractal structure and the difference compared to the square plate is larger at these speeds. This is consistent with the results shown in figure \ref{f13}. 

\begin{figure}
 \begin{center}
   \hspace{-0.5cm} \includegraphics[width=4.13cm]{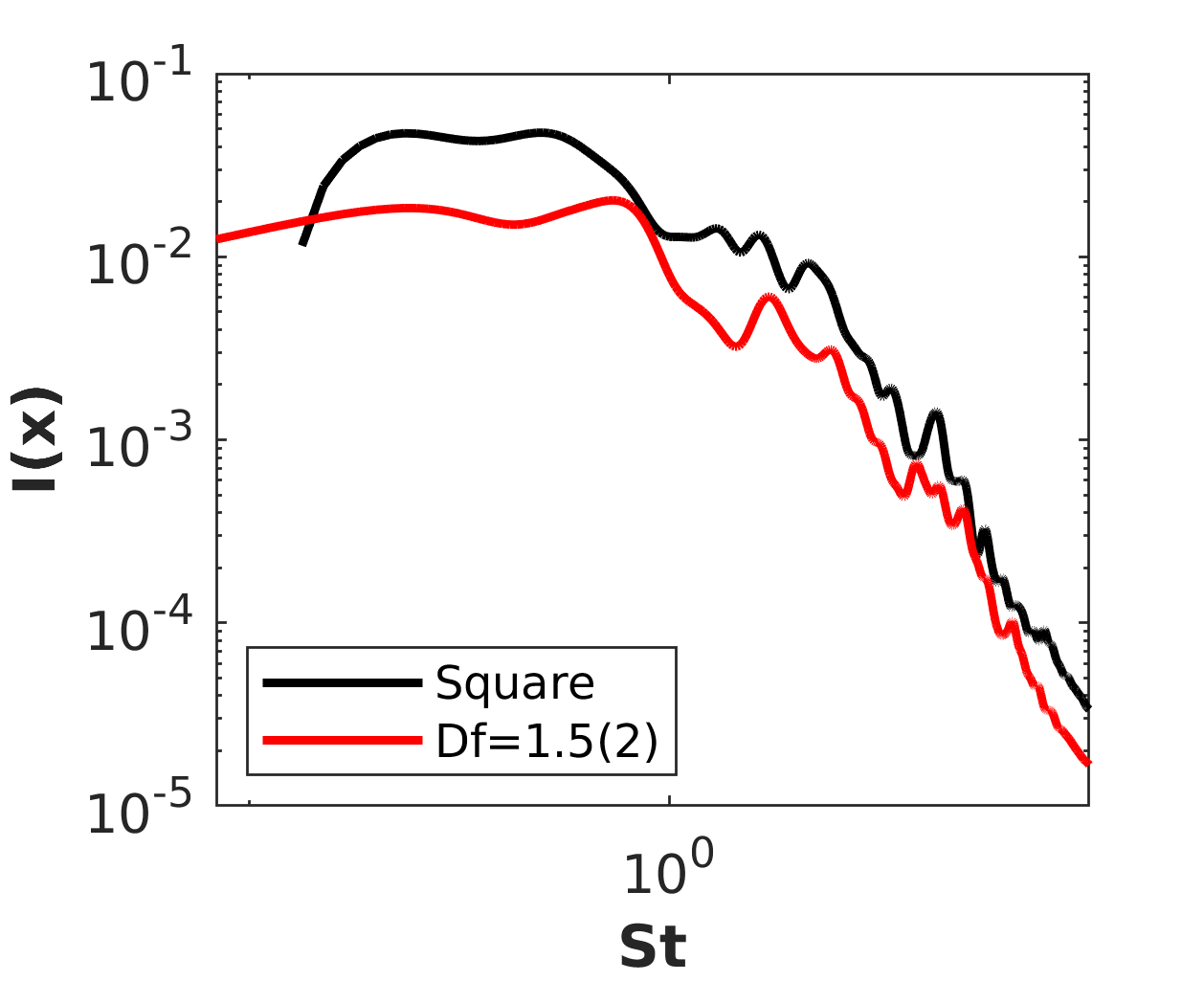}
   \includegraphics[width=4.13cm]{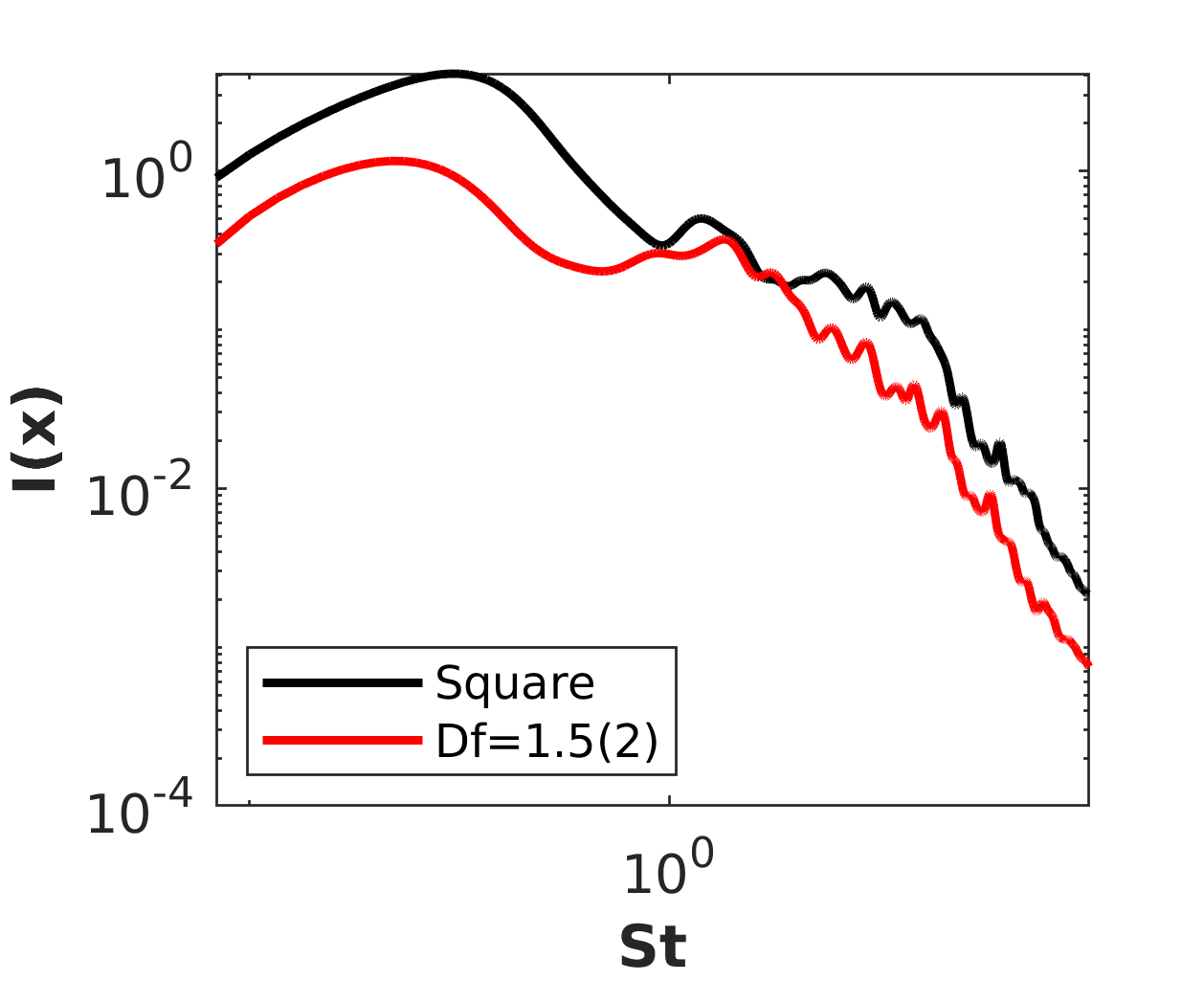} 
   \includegraphics[width=4.13cm]{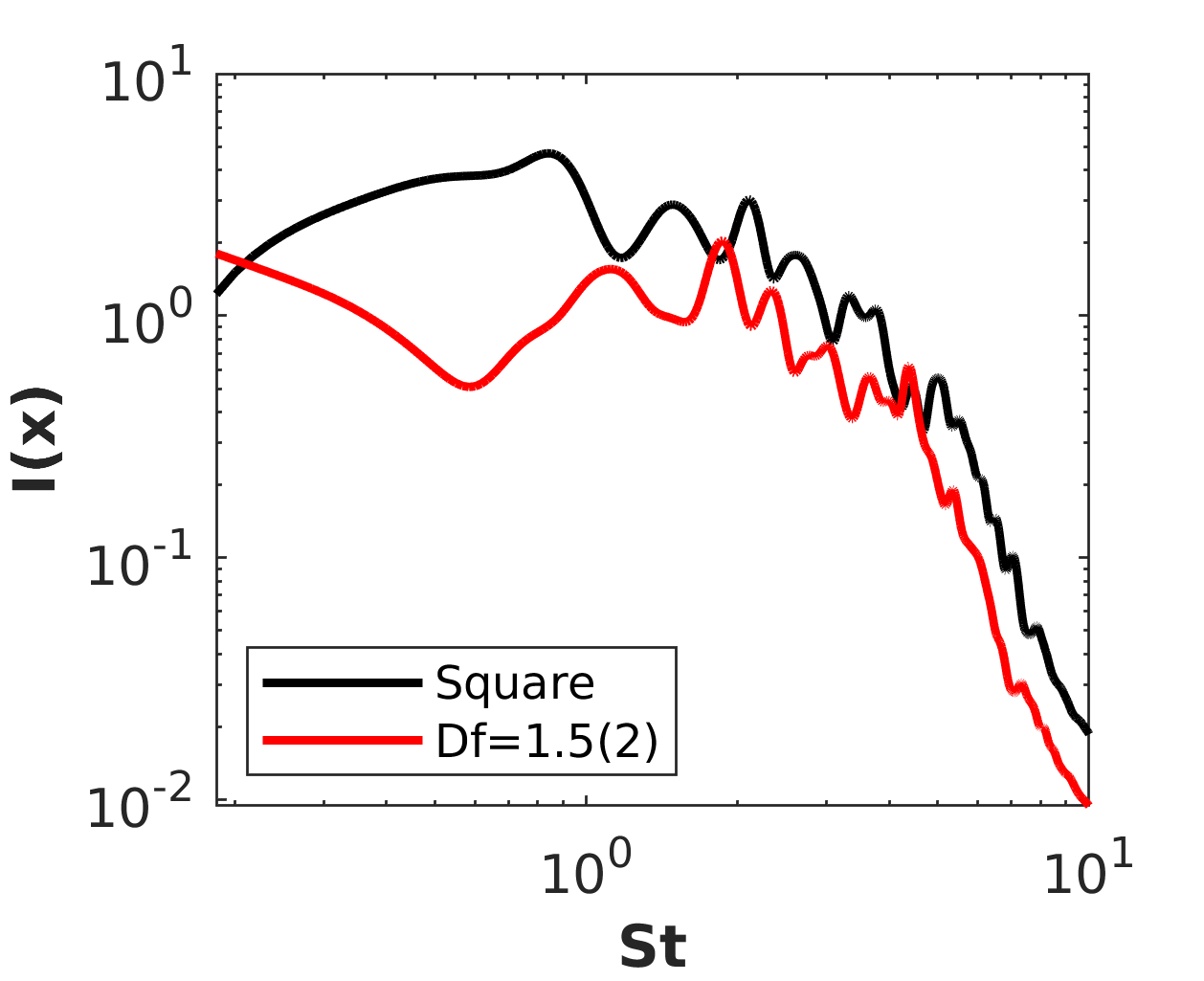}
   \includegraphics[width=4.13cm]{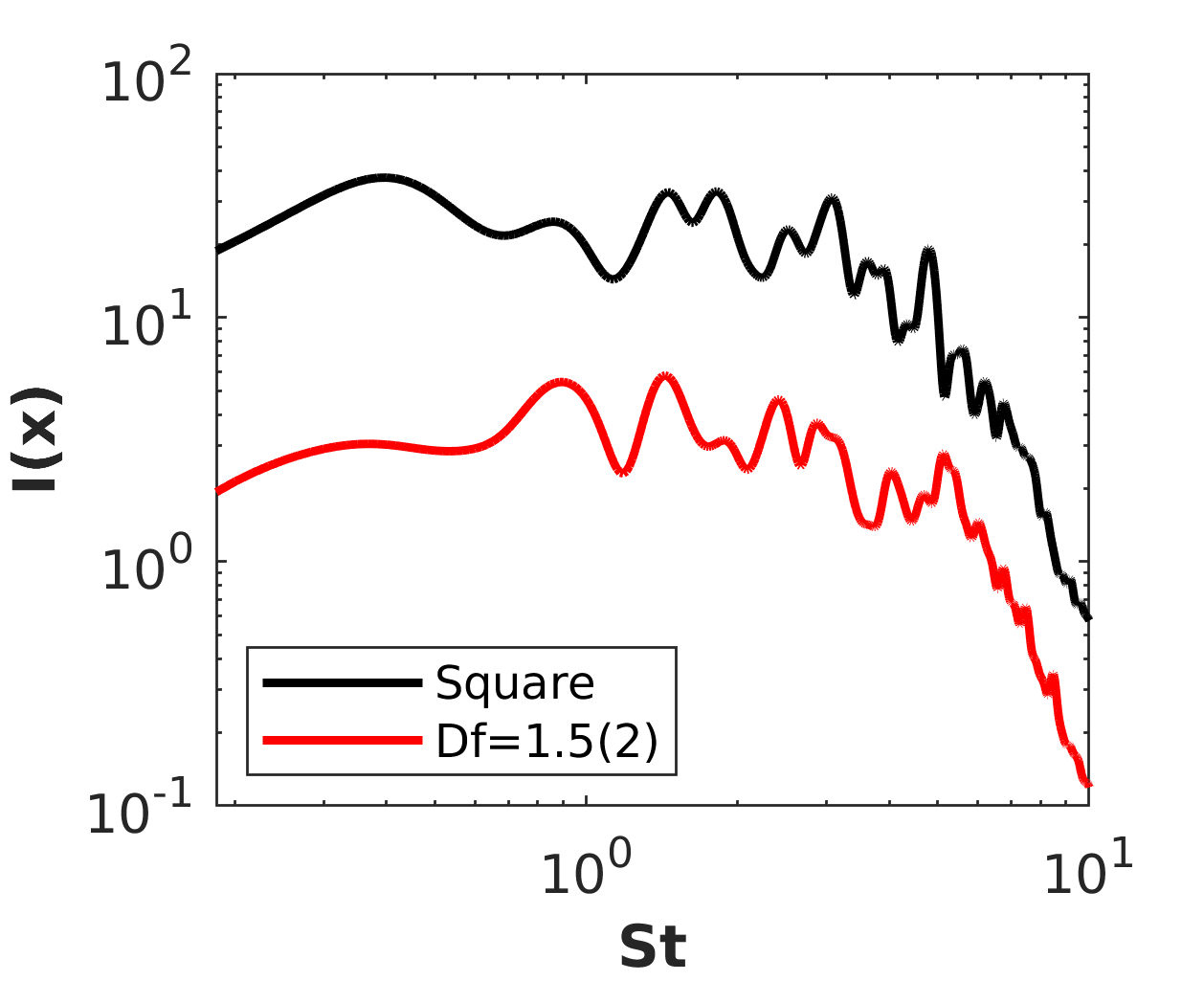}  \\
   (a) \hspace{3.5cm} (b)  \hspace{3.5cm} (c) \hspace{3.5cm} (d)
 \end{center}
  \caption{\label{} Pressure spectrum: a) $M = 0.2$; b) $M = 0.7$; c) $M = 1.4$; d) $M = 2.5$.}
  \label{f17}
\end{figure}

  The noise generated by these flows is expected to be significant, especially for high Mach numbers. To quantify the noise signature, we calculate the pressure spectrum by taking the Fourier transform of the pressure auto-covariance $\langle p'(\mathbf{x},t) p'(\mathbf{x},t+\tau) \rangle$, as
\begin{equation}
 I(\mathbf{x}) = \frac{1}{2\pi} \int_{-\infty}^{\infty} \langle p'(\mathbf{x},t) p'(\mathbf{x},t+\tau) \rangle e^{-i \omega \tau} d \tau
\end{equation}
where $p'$ is the pressure fluctuation, $\tau$ is the time lag, and $\omega$ is the frequency. Figure \ref{f17} shows pressure spectra in $x=4$ just outside of the wake, for the considered Mach numbers; as expected the difference between the results obtained from the square and fractal plates is small for the subsonic cases, and large for the supersonic ones. As illustrated above in figures \ref{f4} and \ref{f5}, the contours show significant pressure fluctuations outside of the wake of the square plate (more significant for the $M=2.5$ case), indicating high level of noise propagating to the far-field. The fractal plate seem to weaken these pressure fluctuations as a subsequent result of breaking up the large flow structures. However, we must mention that this reduction in the pressure spectrum levels in figure \ref{f17}d, for example, is not necessarily an indication of a commensurate far-field noise reduction because these spectra were calculated at a location very close to the wake.

\section{Summary and Conclusions}\label{}

We conducted large eddy simulations of compressible flows interacting with a square and a fractal plate for different Mach numbers. The main objective of the study was to conduct a preliminary investigation of the behavior of high Mach number flows around multi-scale fractal structures, which could potentially mitigate the aerodynamic noise in jet engine applications, or increase mixing desired, for example, in combustion applications. The numerical algorithm is based on high-order low-dissipation spatial and temporal schemes in an implicit LES framework. We modeled the fractal shapes using an immersed boundary method with adequate mesh resolution in regions of small geometrical structures. We validated the numerical algorithm using experimental measurements at a low Mach number, via profiles of the mean flow, Reynolds stresses in the transverse direction, and TKE distribution along the centerline. We found a good agreement between the numerical results and experimental data.

We focused our analysis on a range of Mach numbers that covered both the subsonic and supersonic regimes. We plotted and discussed LES results consisting of iso-surfaces of {\color{black} vorticity magnitude}, contour plots of pressure, mean flow, Reynolds stresses, TKE in the cross-flow or streamwise direction, longitudinal, transverse, and crossflow correlations, as well as velocity and pressure spectra. From the iso-surfaces plots, we noticed some similarities with the incompressible regime in terms of the break up of the flow structures by the fractal plate and the downstream increased mixing. As we increased the Mach number and extended into the supersonic regime, the similarities with the incompressible regime were less noticeable; wake meandering, for example, was not observed in the compressible regime, and the spread of the wake in the lateral direction was smaller. In the supersonic flow cases, contour plots of pressure fluctuations in the near-field revealed larger noise levels in the square plate case when compared to those of the fractal plate, potentially, as a result of the less intense attached shock wave in the case of the fractal plate.

From the mean flow profiles, it was observed that the wake size seemed to decrease with increased Mach number and that the spread rate of the wake is higher in the subsonic regime cases in comparison to those of the supersonic regime. Distributions along the centerline showed the same level of flow reversal for both subsonic cases, while in the supersonic regime cases the level of flow reversal decreased with increasing Mach number. For all Mach numbers, the Reynolds stress profiles indicated that the normal stress components $\langle \rho u_1'u_1' \rangle$ and $\langle \rho u_2'u_2' \rangle$ are of the same order of magnitude for both square and fractal plates, while the shear stress component $\langle \rho u_1'u_2' \rangle$ is at a lower order of magnitude. The fractal plate appeared to reduce the level of all Reynolds stresses when compared to the results obtained from the square plate.

From the two-point velocity correlation plots, the low Mach number cases showed wake meandering in the transverse correlation through negative correlation for large spatial shifts $r$. We also noticed a more rapid de-correlation in the fractal wakes compared to those formed by the square shape. This result not only shows that the fractal rapid de-correlation persists to the compressible regime, but also has important implications for acoustic modeling applications like reducing the peak jet noise. There was not much difference between the correlations corresponding to the square and fractal plates in the supersonic cases, suggesting that the sizes of the flow structures are of the same order of magnitude. We also plotted one-dimensional energy spectra in a point located inside the shear layer and pressure spectra in the near field. The energy spectra showed that for the low Mach number cases, the differences between the results are small for high wavenumbers, but considerably large in the low wavenumber range, which is an indication that the large scale flow structures are broken up by the fractal structures. As we increased the Mach number, the difference between the two spectra became more substantial for all wavenumbers. Pressure spectra showed that there is a potential noise reduction achieved by using fractal plates, and the magnitude of this reduction increases with Mach number.

{\color{black}Our results prove important for extending the findings of previous studies of flows interacting with fractal shapes in the incompressible regime to the compressible counterpart. We showed that fractal-induced behaviors such as mixing, the break up of large flow structures and the corresponding reduction in the aerodynamic sound, which for instance has major practical implications for jet noise, hold in the compressible regime as well, which opens up the door to the implementation of such concepts to more compressible flow engineering applications.}


Future work will aim at extending the analysis to other fractal shapes, and including a freestream turbulence in the upstream. The applicability of these fractal structure to jet noise reduction will also be considered.

\section{Acknowledgements}\label{}

We would like to thank Jovan Nedi$\acute{c}$ from McGill University for providing the experimental data that were used for the validation of the numerical algorithm. M.Z.A. would like to thank Strathclyde University for the financial support from the Chancellor's Fellowship.



\end{document}